\documentclass[usenatbib]{mn2e}
\usepackage{epsfig}
\usepackage{amsmath,amssymb}
\def\lsim{\lower0.6ex\vbox{\hbox{$ \buildrel{\textstyle <}\over{\sim}\ $}}}
\def\gsim{\lower0.6ex\vbox{\hbox{$ \buildrel{\textstyle >}\over{\sim}\ $}}}

\title[The Shape of Galaxy Cluster Haloes]{The Shape of Galaxy Cluster
       Dark Matter Haloes: Systematics of Its Imprint on Cluster Gas,
       and Comparison to Observations}

\author[Flores et al.]
{\parbox[t]\textwidth{Ricardo A. Flores$^1$, Brandon Allgood$^2$,
 Andrey V. Kravtsov$^3$, Joel R. Primack$^4$, David A. Buote$^5$, and
 James S. Bullock$^6$}
\vspace*{6pt} \\
$^1$Department of Physics and Astronomy, University of Missouri --
St. Louis, St. Louis, MO 63121; {\tt ricardo.flores@umsl.edu}
\\
$^2$Physics Department, University of California, Santa Cruz,
CA 95064; {\tt allgood@physics.ucsc.edu}
\\
$^3$Department of Astronomy and Astrophysics, Kavli Institute for Cosmological
Physics, and The Enrico Fermi Institute,\\
$\ $The University of Chicago, Chicago IL 60637;
{\tt andrey@oddjob.uchicago.edu}
\\
$^4$Physics Department, University of California, Santa Cruz,
CA 95064; {\tt joel@scipp.ucsc.edu}
\\
$^5$Department of Physics and Astronomy, University of California,
Irvine, CA 92697; {\tt buote@uci.edu}
\\
$^6$Center for Cosmology, Department of Physics and Astronomy,
University of California, Irvine, CA 92697; {\tt bullock@uci.edu}
\\
}
\date{\today}
\begin{document}
\maketitle

\begin{abstract}
We study predictions for galaxy cluster observables that can test the
statistics of dark matter halo shapes expected in a flat $\Lambda$CDM
universe.  We  present a simple analytical model for the prediction
of cluster--scale X-ray observations, approximating clusters as
isothermal systems in hydrostatic equilibrium, and dark matter haloes
as ellipsoids with uniform axial ratios (homeoidal ellipsoids).  We
test the model against high--resolution, hydrodynamic cluster simulations
to gauge its reliability.  We find that  this  simple prescription
does a good job of predicting cluster X-ray ellipticities compared
to the simulations as long as one focuses on cluster regions that are
less sensitive to recent mergers.  Based on this simple model, the
distribution of cluster--size halo shapes expected in the
concordance $\Lambda$CDM
cosmology implies an X-ray ellipticity distribution with a mean
$\langle \epsilon_X \rangle = 0.32 \pm 0.01$, and a scatter
$\sigma_{\epsilon} = 0.14 \pm 0.01$ for the mass range
$(1-4)\times 10^{14} h^{-1} M_{\odot}$.  We find it important to include
the mass dependence of halo shape when making comparisons to observational
samples that contain many, very massive clusters.  We analyse the systematics
of four observational samples of cluster ellipticities and find that our
results are statistically compatible with observations.  In particular,
we find remarkably good agreement between two recent \textit{ROSAT}
samples and $\Lambda$CDM predictions that \textit{do not} include gas
cooling.  We also test how well our
analytical model can predict Sunyaev--Zel'dovich decrement maps and
find that it is less successful although still useful; the model does
not perform as well as a function of flux level in this case because
of the  changing triaxiality of  dark  matter haloes as  a function  of
radial distance.  Both this  effect  and  the changing  alignment of
isodensity shells of dark matter haloes leave an imprint on cluster gas
that appears to be seen in observational data.  Thus, dark matter haloes
cannot be accurately characterized as homeoidal ellipsoids for all
comparisons.
\end{abstract}


\section{Introduction}

Clusters of galaxies are the largest bound structures in the universe,
and the most recently formed ones according to the very successful cold
dark matter (CDM) cosmology.  As such, their DM haloes are expected to be
less evolved and more aspherical than, say, galaxy-size haloes.  Most gas in
clusters DM haloes has not had time to cool, and since it is gravitationally
subdominant, we can expect it to reflect the underlying 3D shape of
their dark matter haloes.  Indeed, large samples of X-ray clusters
have been known to show a broad distribution of ellipticities in their
surface brightness (SB) maps since the work of \citet*{metal89}.  A
comparison of theoretical predictions to such observations, now that
basic parameters of the underlying cosmology are known at the 10\%--level
or better, may shed light on the basic description of the gas in clusters
of galaxies.

The general expectation that in CDM-based theories DM haloes are flattened,
approximately ellipsoidal, and have short-to-long axial ratios as small
as  $s\equiv c/a \sim 0.5$ has been known for more than 15 years now
\citep{bef87,fwde88}.  Any asphericity in the DM distribution has
important effects on a variety of observed quantities.  In clusters in
particular, asphericity in the dark halo potential will map directly to
asphericity in the gas density, and thus affect the shape of cluster
X-ray isophotes and Sunyaev--Zel'dovich (SZ) isodecrement contours.  Much
subsequent work since these pioneering studies aimed at understanding
the origin of such spatial anisotropy, the influence of the cosmological
model on axial ratios, and improving the resolution with which the
formation of DM haloes was followed \citep{dc91,wetal92,jetal95,tetal98,
suwa03}.  Recently, higher resolution dissipationless simulations have
made it possible {\it to fully characterize the scatter and  mean of axial
ratios}, as a function of both mass and epoch \citep{bu02,js02,ke04,
aetal04}.  \citet[][hereafter JS]{js02}  calculated for the first time
axial ratios for isodensity shells, using cosmological simulations with
$512^3$ particles.  They confirmed that haloes are approximately ellipsoidal
in  isodensity contours, and have provided fits for the dependence of axial
ratios on mass and epoch.  Because  the  isodensity contour method requires
a large number of  particles, JS restricted their analysis to haloes more
massive than $6.2\times10^{12}\ h^{-1}M_\odot\,$ in their $\Lambda$CDM
simulation.  \citet{ke04} have obtained better statistics for haloes more
massive than $3\times10^{14}\ h^{-1}M_\odot\,$ from a Hubble volume
simulation.  We \citep[][hereafter Paper I]{aetal04} have studied haloes
spanning the mass range
$4\times10^{11} - 2\times10^{14} h^{-1} M_{\odot}$ using several
simulations to properly resolve and adequately sample the halo population
in this entire mass range.  Our results are consistent with those of JS
for haloes of low--mass clusters, but yield a steeper mass dependence of axial
ratios than a simple extrapolation of the scaling relations found by JS.
This difference is important in the interpretation of observations on
galaxy scales \citep{aetal04} and, as we show here (Section~\ref{clusters}),
in the interpretation of X-ray ellipticities of samples containing very
massive clusters.  Our results are in agreement with those of
\citet{ke04} if axial ratios are calculated in the same manner.  However,
we find here that axial ratios calculated that way are not useful for
predictions of
cluster observables such as X-ray or SZ maps (see Appendix~\ref{gas}).
A source of uncertainty in the current
understanding of halo shapes is the magnitude of the effect of gas cooling
on cluster DM haloes \citep{ketal04,setal04}.

The variety of current and future observational probes of halo
ellipticity (see Paper I for a discussion) highlights the need to
connect these predictions to observations in a robust fashion.
Here we present an analytic method for predicting gravitational potentials
and cluster gas density based on axial ratios of dark matter haloes.
We test the model against a sample of 8 high--resolution hydrodynamic
ART simulations of 7 clusters (mass $(1-2)\times10^{14} h^{-1} M_{\odot}$)
and 1 group (mass $7\times10^{13} h^{-1} M_{\odot}$) in the $\Lambda$CDM
cosmology \citep{ketal04}, whose highly variant morphology dependence on
the line of sight we exploit to statistically test the model, and 2
additional clusters from earlier
high-resolution simulations (\citealt*{kkh02}; \citealt{nk03}).  We then
apply the model to the prediction of ellipticities for cluster X-ray
and SZ maps.  A perturbative model has been
developed by \citet{ls03}, and further extended by \citet{wf04} to
predict observed distributions from halo shape distributions, but it
is not useful for our purposes here because the ellipticities can be
quite large.

In the following section we discuss the analytic model we use to predict
cluster X-ray and SZ morphologies, and we test it by comparing predictions
for morphology of the simulation clusters (based on their DM haloes only)
with the same observable computed directly from the gas density grid of
the hydrodynamic simulations.  In Section~\ref{clusters} we focus on
observations and compare our predictions to several observational samples
of cluster X-ray ellipticities.  We also discuss recent papers
(e.g. \citealt{floor03}; \citealt*{floor04}) that have compared
observed cluster shapes measured using X-rays and galaxy distributions
to hydrodynamic simulations.  We finish with a summary of our conclusions.
The details of the comparison techniques used in Section~\ref{clusters}
are outlined in three Appendices:
\ref{gas}.~Gas Density Inside Triaxial Halos;
\ref{potential}.~Analytic Potential of Triaxial Generalized NFW Halos;
and \ref{xray}.~A Comparison of X-ray Ellipticity Measures.

\section{Comparison of Model and Simulation Statistics} \label{stats}

In this section we analyse the prediction for two statistics of cluster
morphology, the mean and the dispersion of their ellipticity distribution,
expected in a flat $\Lambda$CDM universe with $\Omega_m = 0.3$, $h  = 0.7$,
and $\sigma_8 =  0.9$.  We first discuss the method to predict cluster
ellipticities based on their dark matter haloes, and then present the
comparison of the predictions to the results from the output of several
high--resolution hydrodynamical simulations \citep{ketal04}.  In what
follows,  the virial radius is defined as the radius, $r_{\rm vir}$,
within which the the mean overdensity drops
to $\Delta = 18\pi^2+82(\Omega_m(z)-1)-39(\Omega_m(z)-1)^2$ \citep{bn98}.
Masses are defined as the mass within $r_{\rm vir}$.

\subsection{Method} \label{method}

We use a diagonalized moment of inertia tensor iteratively calculated
within ellipsoids, or ellipsoidal shells, to define axial ratios for
dark matter haloes \citep{dc91}.  Axial ratios $s=c/a$ and $q=b/a$
($s < q < 1$) are calculated by diagonalizing the tensor
\begin{equation}
M_{ij} \equiv \Sigma \frac{x_i x_j}{R^2}\,;\,\,
R = \sqrt{x^2+y^2/q^2+z^2/s^2}\, ,
\end{equation}
\noindent
thereby determining $s$ and $q$ for the next iteration.
The sum is over all particles within
a given shell $[R,R+\Delta R]$, or the ellipsoid interior to R,
and the iteration starts with
$s = q = 1$.  In Paper I we have found that this method predicts axial
rations that agree with the results of JS, which are based on isodensity
shells, for cluster--size haloes, provided that the axial ratios be
calculated within an ellipsoid of semi-major axis $R = 0.3 r_{\rm vir}$.
Here we also find that the same axial ratios can be used to predict
fairly accurately the mean and dispersion of the expected X-ray
ellipticities, even though as often as half the time the predicted
ellipticity of an individual cluster is off by more than 20\%.

The X-ray surface brightness of an isothermal cluster is given by an
integral over the gas density squared along the line-of-sight (LOS) of
a cluster, SB $\propto \int \rho_{gas}^2 \sqrt{T} \propto \int
\rho_{gas}^2$.  As discussed in Appendix~\ref{gas}, under the assumptions of
hydrostatic equilibrium and isothermality, the gas density at any
point inside a triaxial homeoidal halo can be written in terms of the
temperature $T$, the central gas density, and the halo potential at
the desired point (Eq. \ref{rhogas}).  If we assume that the halo
potential is dominated by the dark matter, then the relation is
simplified by the fact that the potential of any triaxial generalized
NFW halo is analytic (Appendix~\ref{potential}).  Thus, using
the relation (\ref{rhogas}),
we can estimate the X-ray ellipticities implied by a dark matter
halo given the halo axial ratios, and an orientation of a LOS
corresponding to each of the simulation--box axes.

The only (slight) ambiguity in relating the
gas density to the halo potential is the 
the factor $\Gamma$ in equation \ref{rhogas}, which relates
the gas density to the potential exponentially:
$\rho_{gas} \propto \exp(-\Gamma \Phi)$.  
In Appendix~\ref{xray},  we find that the analytic
model works relatively well with $\Gamma \sim c_{\mathit{\rm vir}}\,$
when we compare to two high-resolution clusters. 
This is roughly  expected for an NFW halo since
\begin{equation}
\Gamma \simeq  s q \rho_s R_s^2\frac{G\mu m_p}{k T} 
\simeq \frac{\sigma^{-2} G M_{\mathit{\rm vir}}}
{ r_{\mathit{\rm vir}}\,f(c_{\mathit{\rm vir}})}\,c_{\mathit{\rm vir}}\,
\sim  c_{\mathit{\rm vir}}\,,
\end{equation}
\noindent
where $f(x) = \mathrm{ln}(1+x)-x/(1+x)$ and $\sigma$ is the LOS velocity
dispersion.   In the second step we have used
$s q \rho_s R_s^2 \simeq \rho_{sph} r_s^2$, where $\rho_{sph}$ and $r_s$ are
spherical-NFW-fit parameters for the halo, and assumed the expected energy
scaling, $kT \simeq \mu m_p\sigma^2$, which is even seen observationally
(see e.g. \citealt*{rbn02} and references therein), albeit with a fair
amount of scatter.  The final step follows from rough scaling relations,
and works in detail for the clusters we consider in Appendix~\ref{xray}.
\footnote{For example, for the SCDM ($\Lambda$CDM) cluster discussed
there,
$G\,M_{\mathit{\rm vir}}/r_{\mathit{\rm vir}}\,f(c_{\mathit{\rm
vir}})=1063$ 
($630$) km/s. The dispersion inside the relevant projected radius
($400 \ h^{-1}$ kpc) for these clusters is similar to these values. For the
SCDM cluster, the dispersion is $\sigma=1116$ km/s ($1077$ km/s, $1000$ km/s)
along the x--axis (y--axis, z--axis). For the $\Lambda$CDM cluster,
we find $\sigma=661$ km/s ($928$ km/s, $650$ km/s). The higher $\sigma$
along the y--axis is due to a merger nearly along this axis; however, the
same $\Gamma$ chosen to fit the radial fall--off in surface brightness
in the plane perpendicular to the x--axis works well for the other two
axes.}
Therefore, for our comparisons we assume
\begin{equation}
\frac{GM_{\mathit{\rm vir}}\mu m_p}
{r_{\mathit{\rm vir}}\,f(c_{\mathit{\rm vir}})kT} = 1\,,
\end{equation}
\noindent
and use $\Gamma = c_{\mathit{\rm vir}}\,$. For the dark matter halo of the
SCDM  ($\Lambda$CDM) cluster, we find a  value of $1.06$ ($0.98$) for the
RHS of this equation, using the  average temperature of the gas inside
a radius of $400 \ h^{-1}$ kpc.

\subsection{Results} \label{results}

For a given dark matter halo, the method discussed above allows us to
compute the SB map expected for a given LOS through that halo.  We
discuss in Appendix~\ref{xray} how an X-ray ellipticity can the be
obtained from the SB map.  There is no unique method to calculate
ellipticities and, as we discuss below and in Appendix~\ref{xray}, it
is important to follow the procedure chosen by observers to calculate
ellipticities in order to compare to observations.  {\it Individual}
ellipticities can differ {\it substantially} depending on what part of
a map the procedure selects and, as we show below, {\it even the means}
(of samples of ellipticities calculated with different procedures)
{\it will differ}.

Figure~\ref{allCL.SBmaps} shows SB maps for 8 high--resolution
{\it adiabatic} hydrodynamic simulations  of clusters in the
$\Lambda$CDM cosmology \citep[see][]{ketal04}.
Each row in the figure shows the 3 SB maps corresponding to a LOS parallel
to each of the coordinate axes of the simulation box containing the cluster.
We calculate the SB for a given ``pixel'' in each box by summing
$\rho_{gas}^2\sqrt{T}$ over all cells along the LOS--axis. Each cell is
$7.8 \ h^{-1}$ kpc on the side, and each map covers $2 \ h^{-1}$ Mpc on the
side.  The X-ray ellipticity, $\epsilon_X$, shown in the upper right
corner of each map, is calculated using the pixels (shown by the shaded
areas) containing 20\% of the
total flux above a threshold flux that is 1\% of the peak flux of the map.
This is one of the procedures we consider in this work to calculate
ellipticities from a SB map.  It is a method that in the absence of noise
yields ellipticities that reflect the potential of the DM halo, as
we show in the next paragraph.  \citet{metal89} used this method in their
study of X-ray ellipticities.  However, we find below (see
Section~\ref{clusters}) that their data are heavily affected by noise
and do not serve as a test of the $\Lambda$CDM cosmology.
A more detailed discussion of methodologies is presented in
Appendix~\ref{xray}.

For each cluster halo we can use the method described in Section~\ref{method}
to compute the predicted SB map for a given LOS through a given cluster.  We
can then compute the predicted X-ray ellipticity in exactly the same manner
we computed the ellipticity for the hydrodynamic simulation maps of
Figure~\ref{allCL.SBmaps}. In Figure~\ref{HydroVSmodel} we show a comparison
of the two ellipticities.  Each point in the figure plots the ellipticity
shown in Figure~\ref{allCL.SBmaps} against the ellipticity calculated based
only on the DM halo parameters.  A Kolmogorov--Smirnov (KS) test gives
high probability ($P_{KS} = 99\%$) that the two sets represent the same
distribution.  We treat the value of $\epsilon_X$ for each LOS as an
independent measurement because for a given axial ratio $s$ there is quite
a degree of variability expected for the other axial ratio, and then there
is the variation introduced by the orientation of the cluster to the LOS.
There is indeed quite a degree of variation in $\epsilon_X$ for each cluster,
but a stronger test will only be possible with many more simulations.
The means of the sets indeed agree quite well:
$\langle \epsilon_X^{hydro} \rangle = 0.36$, whereas
$\langle \epsilon_X^{model} \rangle = 0.35$.  However, the dispersions
differ significantly: $\sigma_{\epsilon}^{hydro} = 0.18$, whereas
$\sigma_{\epsilon}^{model} = 0.12$.  This is due to the middle map in the
last row of Figure~\ref{allCL.SBmaps},
whose ellipticity is greatly enhanced by
a secondary lump that in this case has a significant relative weight due
to fact that the flux levels select a narrow region of the main cluster.
Without that map, $\langle \epsilon_X^{hydro} \rangle = 0.34$, and
$\sigma_{\epsilon}^{hydro} = 0.15$ (recalculating the ellipticity without
the lump, $\langle \epsilon_X^{hydro} \rangle = 0.35$ and
$\sigma_{\epsilon}^{hydro} = 0.15$).  The remaining difference between
$\sigma_{\epsilon}^{hydro} = 0.15$ and $\sigma_{\epsilon}^{model} = 0.12$
seems to be due to the fact that
the gas reflects the changing triaxiality in the inner region of DM
haloes.  We tested this by recalculating $\epsilon_X^{model}$ using axial
ratios for the DM haloes calculated within $R = 0.15 r_{\rm vir}$, in
which case we find $\langle \epsilon_X^{model} \rangle = 0.35$ and
$\sigma_{\epsilon}^{model} = 0.15$.

%%%%%%%%%%%%%%%%%%%%%%%%%%%        Figure1        %%%%%%%%%%%%%%%%%%%%%%%
\begin{figure*}
\epsfig{file=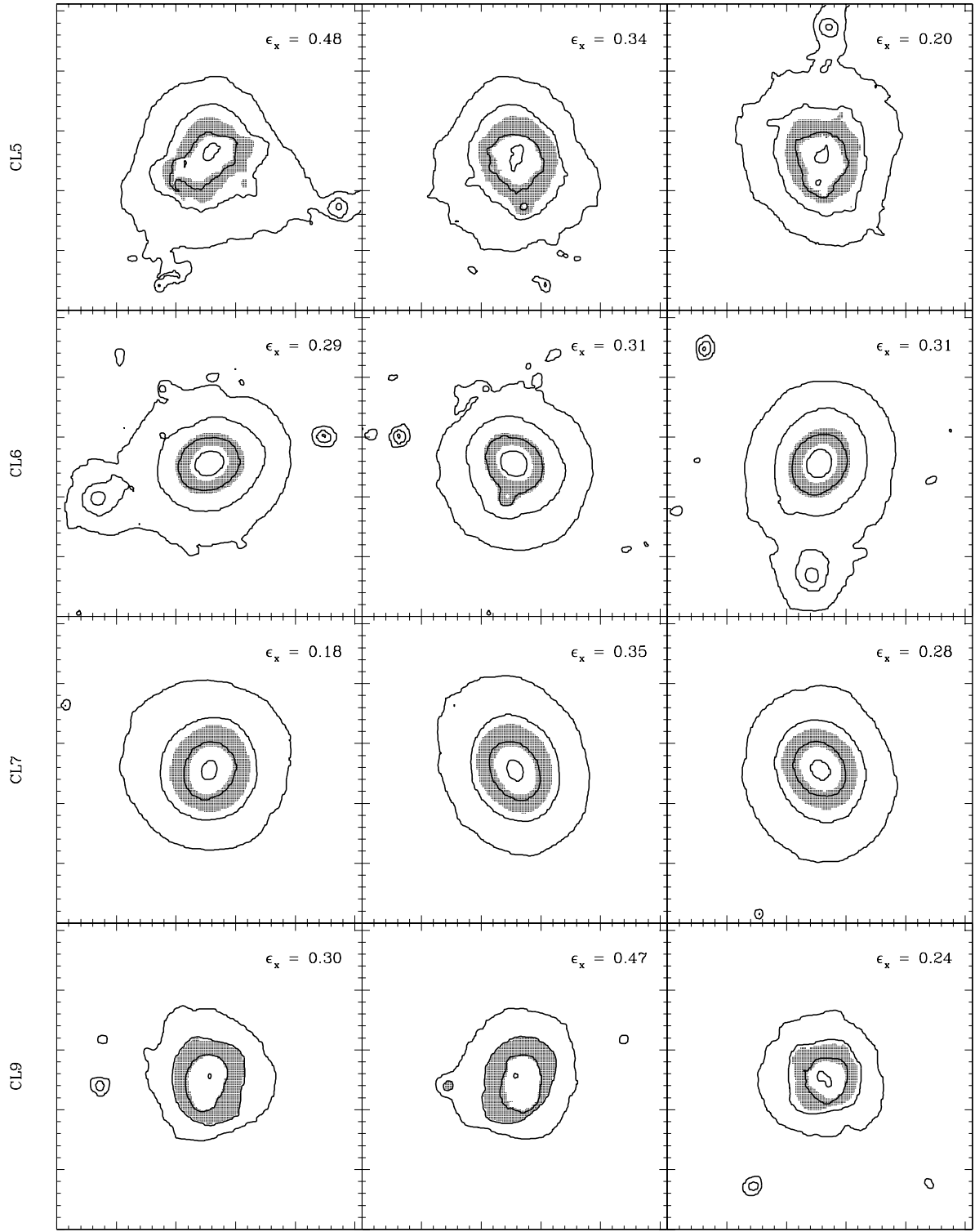,width=\hsize}
\caption{X-ray SB maps for hydrodynamic simulation clusters.  Each row
presents the SB for a LOS along each of the axes of the simulation box.
Each square is $2 \ h^{-1}$ Mpc across. The solid lines show contours
of constant SB, spaced by factors of 10.  The shaded area is the region
used to calculate the ellipticity shown in the upper right corners. See
text for explanation and discussion.
\label{allCL.SBmaps}}
\end{figure*}

\addtocounter{figure}{-1}

\begin{figure*}
\epsfig{file=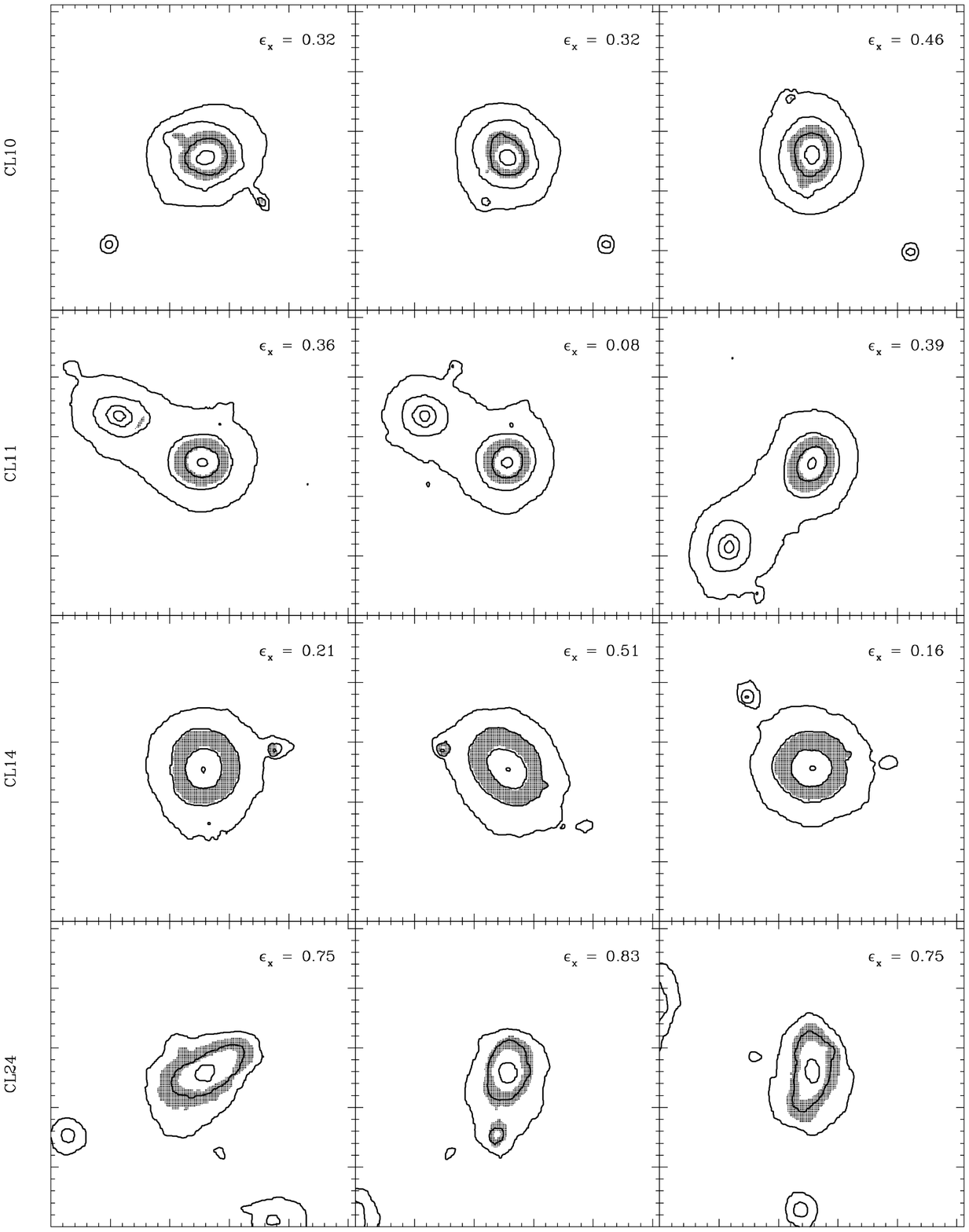,width=\hsize}
\caption{Continued}
\end{figure*}
%%%%%%%%%%%%%%%%%%%%%%%%%%%%%%%%%%%%%%%%%%%%%%%%%%%%%%%%%%%%%%%%%%%%%%%%%

%%%%%%%%%%%%%%%%%%%%%%%%%%%        Figure2        %%%%%%%%%%%%%%%%%%%%%%%
\begin{figure}
\epsfig{file=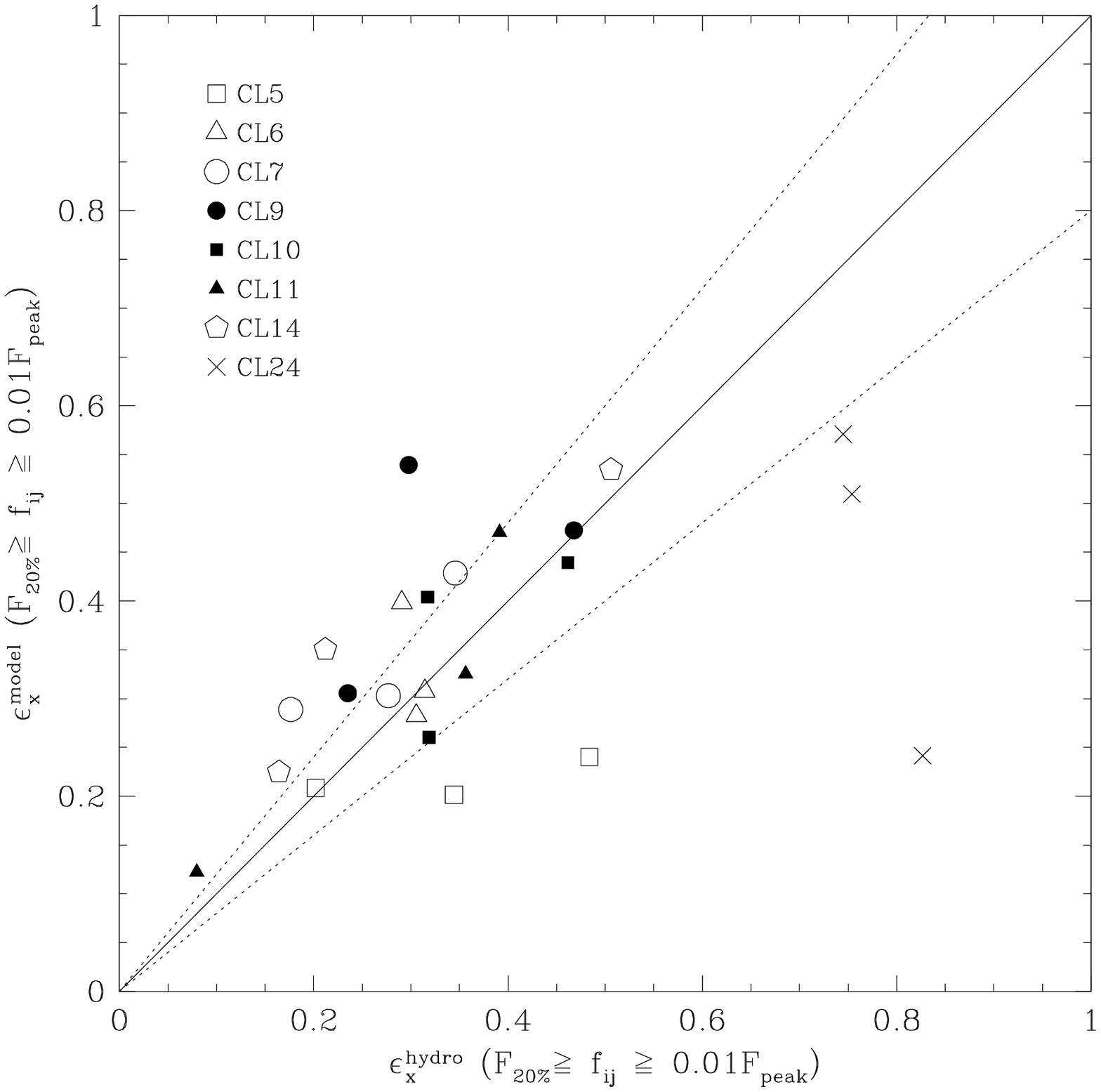,width=0.99\hsize}
\caption{Model X-ray ellipticities compared to the ellipticities calculated
directly from SB maps for the hydrodynamic simulation clusters.  Each
symbol identifies a cluster, and for each cluster the symbol plots the
ellipticity calculated from the simulation map (one for each LOS along each
of the coordinate axes) against the predicted ellipticity using the method
described in Section~\ref{method}.  Within the dotted lines the
ellipticities differ by less than 20\%.
\label{HydroVSmodel}}
\end{figure}
%%%%%%%%%%%%%%%%%%%%%%%%%%%%%%%%%%%%%%%%%%%%%%%%%%%%%%%%%%%%%%%%%%%%%%%%%

%%%%%%%%%%%%%%%%%%%%%%%%%%%        Figure3        %%%%%%%%%%%%%%%%%%%%%%%
\begin{figure}
\epsfig{file=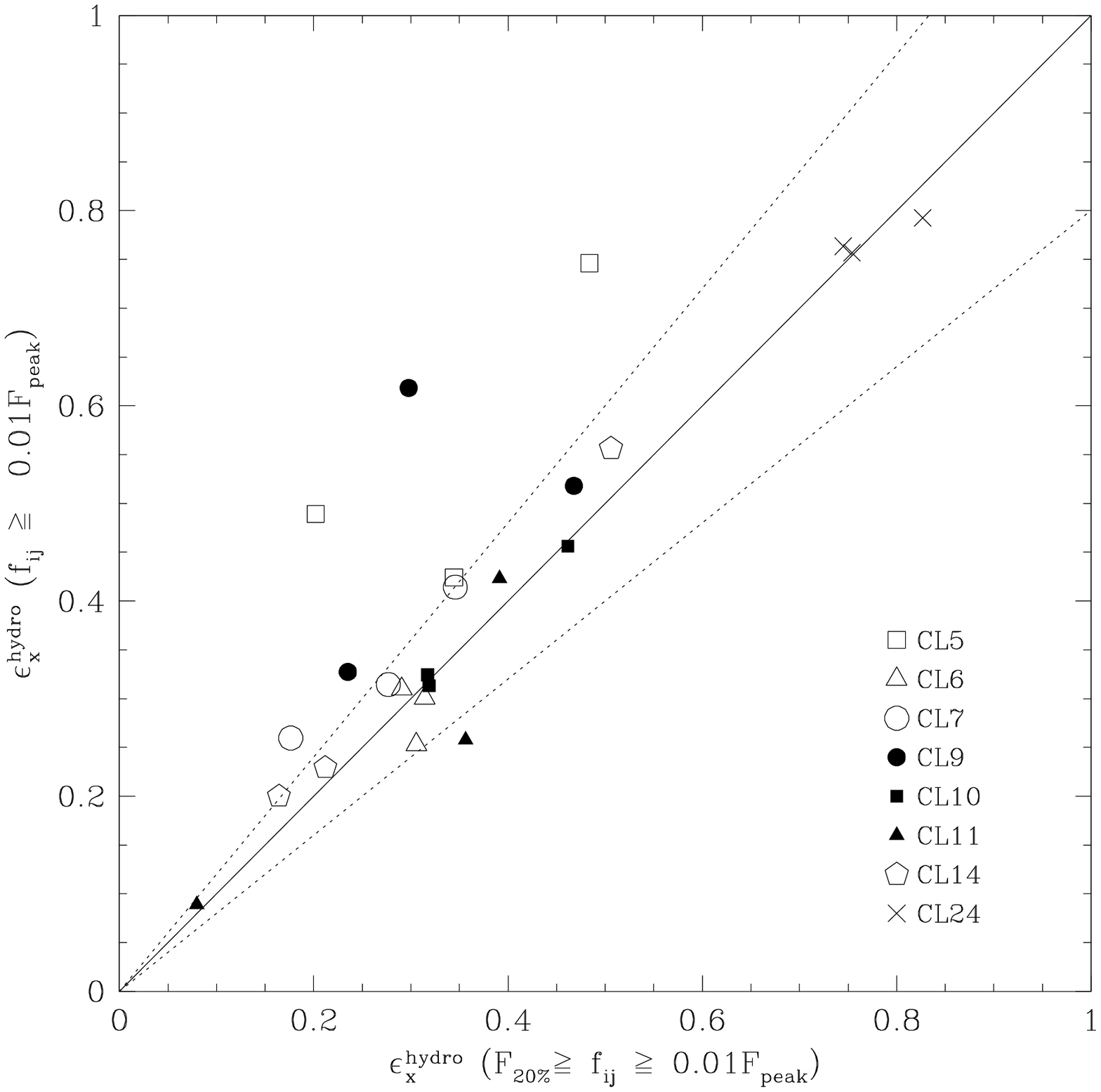,width=0.99\hsize}
\caption{Comparison of ellipticities calculated directly from SB maps of
the hydrodynamic simulation clusters using different strategies.  The
abscissa is the same as in Figure~\ref{HydroVSmodel}.  The ordinate is
an ellipticity calculated using {\it all} pixels above a given flux
threshold, in this case 1\% of the peak flux in a map.  Within the dotted
lines the ellipticities differ by less than 20\%.  See text for further
discussion.
\label{strategies}}
\end{figure}
%%%%%%%%%%%%%%%%%%%%%%%%%%%%%%%%%%%%%%%%%%%%%%%%%%%%%%%%%%%%%%%%%%%%%%%%%

%%%%%%%%%%%%%%%%%%%%%%%%%%%        Figure4        %%%%%%%%%%%%%%%%%%%%%%%
\begin{figure}
\epsfig{file=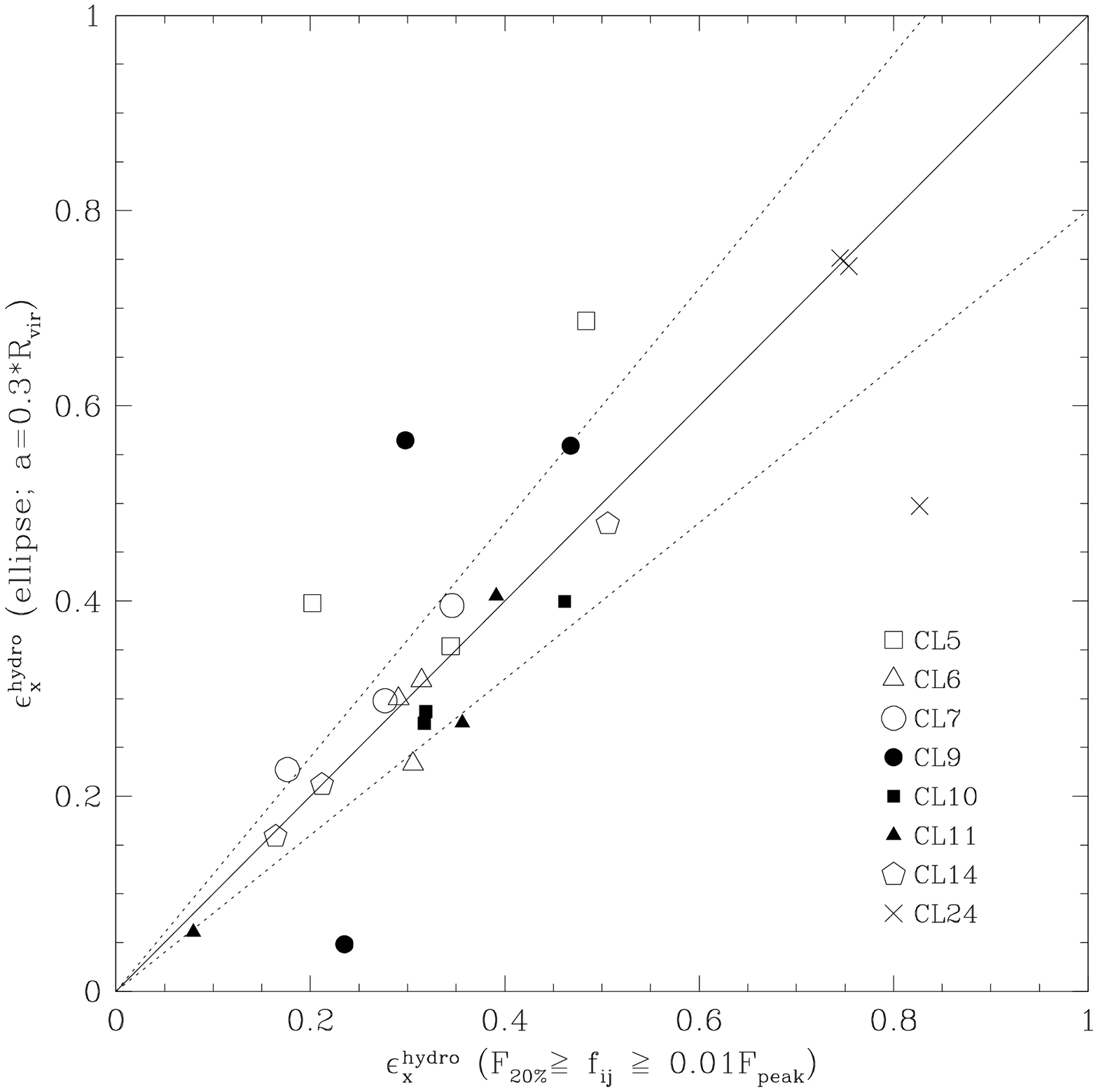,width=0.99\hsize}
\caption{Comparison of ellipticities calculated directly from SB maps of
the hydrodynamic simulation clusters using different strategies.  The
abscissa is the same as in Figure~\ref{HydroVSmodel}.  The ordinate is
an ellipticity calculated iteratively using {\it all} pixels within an
elliptical aperture of fixed semi-major axis ${\rm a} = 0.3 r_{\rm vir}$.
Within the dotted lines the ellipticities differ by less than 20\%.  See
text for further discussion.
\label{buote}}
\end{figure}
%%%%%%%%%%%%%%%%%%%%%%%%%%%%%%%%%%%%%%%%%%%%%%%%%%%%%%%%%%%%%%%%%%%%%%%%%

%%%%%%%%%%%%%%%%%%%%%%%%%%%        Figure5        %%%%%%%%%%%%%%%%%%%%%%%
\begin{figure*}
\epsfig{file=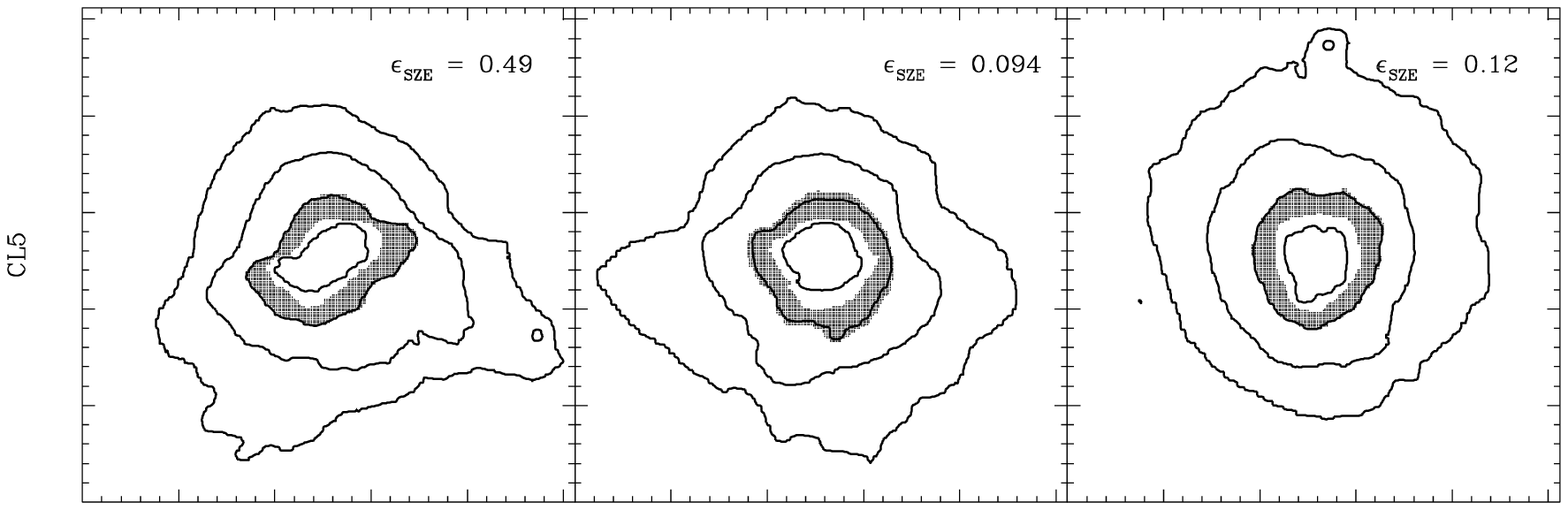,width=\hsize}
\caption{Sunyaev--Zel'dovich decrement maps for a hydrodynamic simulation
cluster.  Each panel presents the map for a LOS along each of the axes
of the simulation box.  Each square is $2 \ h^{-1}$ Mpc across. The solid
lines show contours of constant temperature decrement, spaced by factors
of 3.  The shaded area is the region used to calculate the ellipticity
shown in the upper right corners. See text for explanation and discussion.
\label{allCL.SZEmaps}}
\end{figure*}
%%%%%%%%%%%%%%%%%%%%%%%%%%%%%%%%%%%%%%%%%%%%%%%%%%%%%%%%%%%%%%%%%%%%%%%%%

Thus, the analytic model can be
used to make fairly robust predictions of average X-ray ellipticities.
An equally robust prediction of the expected scatter does not seem
possible with a simple homeoidal model of DM haloes, but its reliability
might be checked by calculating two sets of ellipticities based on DM
halo axial ratios calculated within two different radii.
We have used the model, then, to calculate the mean and dispersion of 
X-ray ellipticities expected in a $\Lambda$CDM universe
at the present epoch.  We use a sample of 46 DM haloes extracted from the
$120 \ h^{-1}$ Mpc dissipationless cosmological simulation discussed
in Paper I.  The cosmology is a flat $\Lambda$CDM universe with
$\Omega_m = 0.3$, $h  = 0.7$,  and $\sigma_8 =  0.9$ and the simulation
followed $512^3$ particles of mass $1.1\times10^{9}\ h^{-1}M_\odot$.
All haloes with virial mass $(1-4)\times 10^{14}  h^{-1} M_{\odot}$ were
selected.  We calculate their axial ratios and concentrations in order
to predict X-ray ellipticities for a LOS corresponding to each of the
coordinate axes of the box.  The ellipticity is computed as described
above, using two flux levels.  The samples corresponding to each LOS
agree quite well with each other.  For the combined sample we find
\begin{equation}
\label{prediction}
\langle \epsilon_X \rangle = 0.323 \pm 0.013\,;\,\,
\sigma_{\epsilon} = 0.138 \pm 0.008\, ,
\end{equation}
\noindent
where the errors are calculated by bootstrap resampling.  These results
are consistent with those for the hydrodynamic simulation clusters, for
which $\langle \epsilon_X \rangle = 0.338 \pm 0.032$ and
$\sigma_{\epsilon} = 0.148 \pm 0.030$.

The strategy to extract an X-ray ellipticity from a SB map is by no
means unique, and in Figure~\ref{strategies} \& Figure~\ref{buote} we
present two other cases of interest here.
For example, in Section~\ref{clusters} we discuss a sample
of X-ray ellipticities obtained by \citet{ketal01} who use a strategy
that emphasizes the central region of a cluster (they were interested in
mergers).  In Figure~\ref{strategies} we show a comparison of ellipticities
(all calculated directly from the hydrodynamic simulation maps) using two
different strategies.  The ordinate is an ellipticity very similar to that
of \citet{ketal01}, calculated using all pixels above a flux threshold
corresponding to 1\% of the peak flux of the SB map.  The abscissa is as
in Figure~\ref{HydroVSmodel}.  It can be seen there that they differ
systematically from one another: the means differ by 14\%.  Therefore,
a direct comparison of a sample of ellipticities calculated in this
fashion to our predictions, (Eq. \ref{prediction}), is not possible.

Another example of interest here is the strategy used by \citet*{betal05}.
They calculate ellipticities using {\it all} pixels inside a smooth boundary
(i.e. the boundary is not determined by flux level), which is determined
by applying the method of \citet{cm80} (used without iteration in the
studies of \citealt{metal89} and \citealt{ketal01} as explained in
Appendix~\ref{xray}) iteratively, starting from a circle, until the
ellipticity converges with a given accuracy. The semi-major axis is
kept fixed.  Figure~\ref{buote} shows a comparison of ellipticities
(all calculated directly from the hydrodynamic simulation maps), calculated
using the methodology of \citet{betal05} (ordinate) and \citet{metal89}
(abscissa).  The ellipticities agree quite well in mean value and
dispersion ($P_{KS} = 89\%$), despite the fact that the methodology of
\citet{betal05} uses all pixels within the elliptical window.  Thus, the
choice of a smooth boundary (rather than a flux--selected boundary) makes
the ellipticity samples differ in no systematic way, unlike the case of
Figure~\ref{strategies}.  A comparison of a sample of ellipticities
calculated this way to our predictions, (Eq. \ref{prediction}), is
therefore possible.

Finally, we also explore here the reliability of the analytic model to
predict the expected ellipticity of millimeter--wave maps of the SZ effect
(SZE) in clusters \citep[see e.g.][]{chr02}, which map the effective
temperature decrement of the microwave background due to the hot
electron gas \citep{sz70}.
Figure~\ref{allCL.SZEmaps} shows decrement maps for one of the clusters
in Figure~\ref{allCL.SBmaps} (CL5), with contours spaced by a factor of
3 (the maps for all clusters are included as supplementary material).
The maps are qualitatively similar to
the SB maps, but the effect of changing triaxiality of the DM haloes in
the region spanned by the isodecrement contours shown is more readily
noticed (because the signal is proportional $\rho_{\rm gas}$ instead of
$\rho_{\rm gas}^2$).  We show in the
upper right corner of each map the ellipticity
obtained in the same manner as Figure~\ref{allCL.SBmaps}, but the
decrement threshold and the percentage of signal in the pixels are
chosen so that the pixels used cover a region of similar size to the
corresponding region in Figure~\ref{allCL.SBmaps}.  Specifically, the
decrement threshold chosen is 10\% of the peak decrement in the map
(as opposed to the 1\% of peak signal in Figure~\ref{allCL.SBmaps}), and
the signal in all of the pixels used is 30\% of the total signal above
the threshold (as opposed to the 20\% in Figure~\ref{allCL.SBmaps}).
The mean ellipticity and the scatter for this set are
$\langle \epsilon_{SZE} \rangle = 0.307 \pm 0.035$ and
$\sigma_{SZE} = 0.171 \pm 0.019$, respectively (bootstrap resampling
errors).  For the set of Figure~\ref{allCL.SBmaps}, but with 10\% threshold
and 30\% flux, $\langle \epsilon_X \rangle = 0.359 \pm 0.036$ and
$\sigma_{X} = 0.175 \pm 0.033$, respectively (the difference with our
1\% threshold and 20\% flux prediction above is due to the changing
triaxiality of DM halos).  By contrast, the analytic
model would predict nearly identical destributions.  Thus, although not
as successful as for X-ray ellipticities, the analytic model would still
be useful to predict, e.g. quantitative trends for ellipticities as a
function of cluster redshift.

We now turn to a quantitative comparison of the expected distribution of
X-ray ellipticities in the $\Lambda$CDM cosmology, obtained using the
analytic model we have discussed and tested here, with observational
samples.

\section{Comparison to Cluster--scale Observations} \label{clusters}

Here we compare our predictions to ellipticity distributions from samples
of cluster X-ray observations.  We first analyse the methodology
employed by \citet{metal89} and \citet{ketal01} to calculate ellipticities
for their samples of Abell clusters.  We also consider briefly
the sample of \citet{metal95} considered by \citet{wf04} for their
comparison to observations.  These samples use different methodology to
calculate ellipticities, and are affected differently by resolution and
noise.  Applying a KS test to pairs of samples (all converted to 2D axial
ratios) we find that $P_{KS} = 0.0031$ ($P_{KS} = 0.21$) for the
\citet{metal89} and \citet{metal95} samples (\citet{ketal01} and
\citet{metal95} samples).  This complicates the comparison of theoretical
predictions and observations, but it is usually ignored
\citep[e.g.][]{melott01,wf04}.  Finally, we analyse a very recent data
set from a nearly complete, flux--limited sample of \textit{ROSAT} clusters
discussed by \citet{betal05}.

\citet{metal95} considered a sample of 51 (mostly Abell) clusters observed
by the {\it Einstein} IPC, for
which they obtained a mean 2D axial ratio, $\eta$, of
$\langle \eta \rangle = 0.80$ and a dispersion $\sigma_{\eta} = 0.12$.
Converting their axial ratios to ellipticities, $\epsilon = 1 - \eta^2$,
we obtain $\langle \epsilon \rangle = 0.358 \pm 0.026$ and
$\sigma_{\epsilon} = 0.182 \pm 0.017$.  The mean and scatter differ by
about 1.5 and 2.5 standard deviations, respectively, from our predictions
(Eq. \ref{prediction}).  However, the method of \citet{metal95} uses all
pixels above a S/N level, and therefore gives substantially more weight
to the central regions of a SB map, where mergers can significantly
affect the ellipticity.  Given our discussion of the results presented in
Figure~\ref{strategies}, the difference in mean ellipticity (10\%) is
entirely within the expectection given the different strategy.  The
agreement is somewhat surprising, however, given the potential effect
that cooling within clusters could have on the DM haloes
\citep{ketal04,setal04}.  We discuss this further below.

The \citet{ketal01} sample consists of 22 \textit{ROSAT} clusters, with a
range of velocity dispersions of $400-1000$ km/s.
Converting their ellipticities to $\epsilon = 1 - \eta^2$, the mean and
dispersion of their sample are $\langle \epsilon \rangle = 0.458 \pm 0.051$
and $\sigma_{\epsilon} = 0.237 \pm 0.023$.
The poor agreement with our prediction (Eq. \ref{prediction}) is not
surprising given that their method emphasizes the cluster centre and
there are three clusters
in the observational sample showing strong evidence of an
ongoing merger: A2804, A2933, and A3128 are all bimodal \citep{ketal01}.
We have tested that this is indeed the problem by computing ellipticities
for the sample of hydro clusters discussed in Section~\ref{stats},
following exactly the procedure of \citet{ketal01}, which first defines
a flux threshold equal to the mean flux within a $600 \ h^{-1}$ kpc
radius, and then uses all pixels above the threshold.  A KS test between
the hydro sample of ellipticities calculated this way, and the sample
of \citet{ketal01}, gives $P_{KS} = 0.82$ (the hydro sample is slightly
rounder on average).  Thus we conclude that their sample is in agreement
with the expectations for a $\Lambda$CDM universe.

We have also made a comparison with the~\textit{Einstein} data of
\citet{metal89} consisting of 49 clusters.  Here we can expect the
comparison to be a better test on the cosmological sample because they
explicitly exclude image centres, thus their shape statistic is less
sensitive to mergers (see Appendix~\ref{xray}).  However, the mean and
dispersion of their sample are
$\langle \epsilon \rangle = 0.240 \pm 0.020$
and $\sigma_{\epsilon} = 0.142 \pm 0.015$.
There is poor agreement with our prediction (Eq. \ref{prediction}) for
the mean this time.
It appears unlikely that this discrepancy could be entirely due to missing
physics (e.g. cooling) in the simulations we have used to test the analytic
model described in Appendix~\ref{gas}.  We note that even after excluding
the 3 bimodal clusters from the \citet{ketal01} sample, a KS test against
the \citet{metal89} sample (once ellipticities are converted to the same
definition in terms of flux--moment eigenvalues; see Appendix~\ref{xray})
rejects that they are compatible at the 96\% CL.

%%%%%%%%%%%%%%%%%%%%%%%%%%%        Figure6        %%%%%%%%%%%%%%%%%%%%%%%
\begin{figure}
\epsfig{file=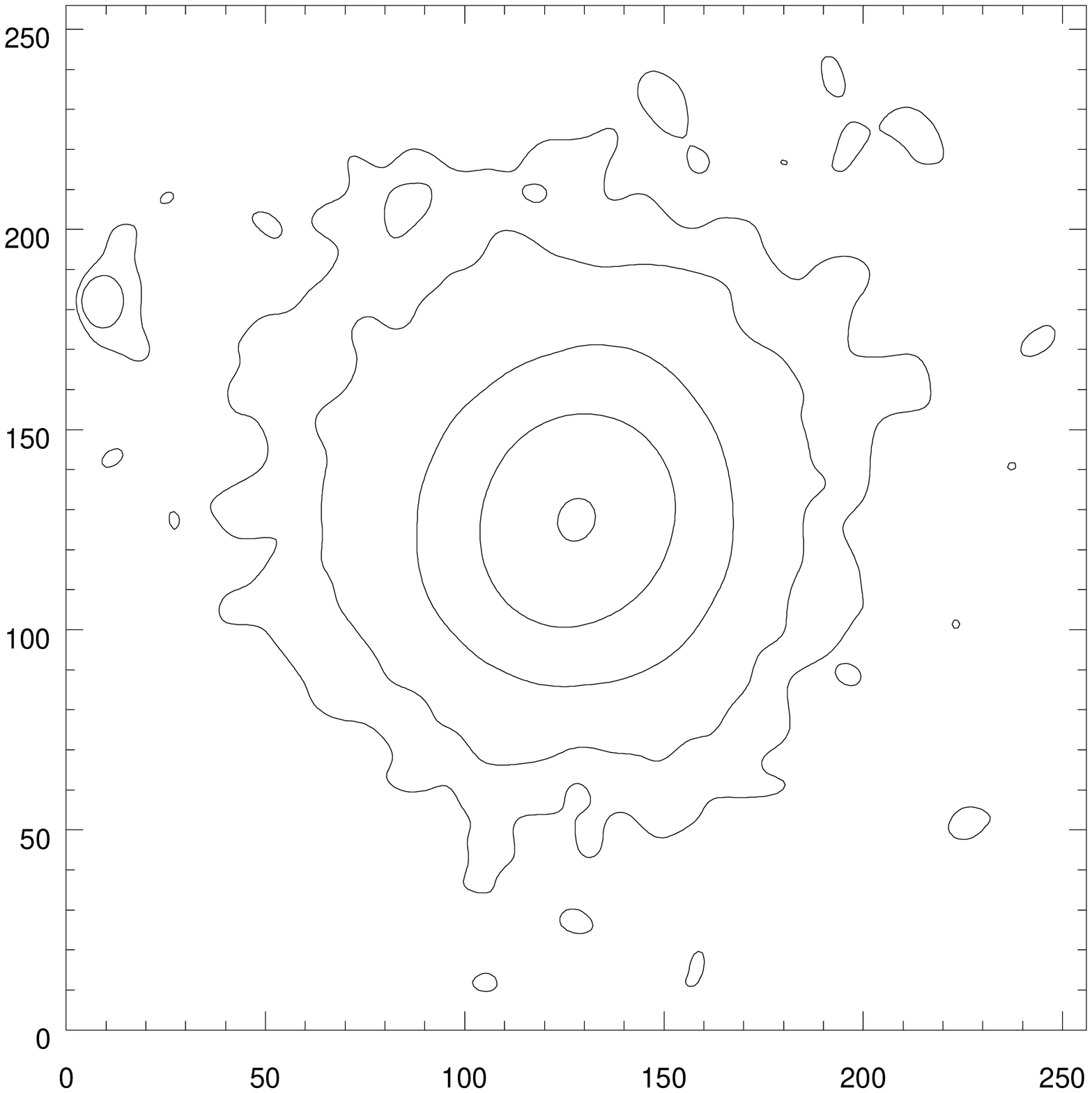,width=0.99\hsize}
\caption{Simulated X-ray map of a hydrodynamic simulation cluster (CL7)
after smoothing and adding noise.  See text for discussion.
\label{noise}}
\end{figure}
%%%%%%%%%%%%%%%%%%%%%%%%%%%%%%%%%%%%%%%%%%%%%%%%%%%%%%%%%%%%%%%%%%%%%%%%%

The coarser angular resolution of the~\textit{Einstein} data probably
contributes to this disagreement.
For example, we find that if we smooth the X-ray map of the $\Lambda$CDM
cluster we discuss in Appendix~\ref{xray} (see Figure~\ref{LCDMSBmap})
with a gaussian window of $80 \ h^{-1}$ kpc (FWHM, roughly corresponding
to the $1.6\arcmin$ resolution of the \citet{metal89} data at the median
redshift of their sample, $z = 0.057$), the X-ray ellipticities can change
significantly: the entry in column 6 of Table~\ref{tbl-1} would be 0.40
(0.23, 0.46) for the x--axis (y--axis, z--axis) as compared to 0.43
(0.28, 0.51) without smoothing.  This is also consistent with the changes
found by \citet{bc96} for 5 clusters with \textit{Einstein}, later analysed
with \textit{ROSAT} data.  We find that a 20\% change in the predicted
ellipticities would make them marginally compatible with the data.

A more important contribution to the difference with our predictions
seems to be the effect of noise.  Many clusters in the \citet{metal89}
sample have very small ellipticities but do not look round at all.
Figure~\ref{noise} shows the effect of smoothing and noise on one of the
clusters in Figure~\ref{allCL.SBmaps} (CL7, left).  The noise level is
at 70\% of the threshold (1\% of peak flux) used for the calculation of
$\epsilon_X$.  The automated software
of \citet{metal89} allowed up to 50 islands in the X-ray map, thus this
map would be accepted.  The ellipticity without smoothing and noise
would be $\epsilon_X = 0.18$ in this case, but with smoothing and noise
(the latter making the largest difference) $\epsilon_X = 0.08$.
If we put this level of noise in all the hydro clusters, we find that
the hydro sample becomes fully compatible with the \citet{metal89}
sample: $P_{KS} = 0.87$.  Thus, it may be that our predictions are not
incompatible with this data sample.

Finally, we discuss the recent sample of ellipticities obtained by
\citet{betal05} for the flux--limited sample of \textit{ROSAT} clusters
of \citet{randb02}. As we showed in Section~\ref{results}, a direct
comparison to our predictions (Eq. \ref{prediction}) is in this case
possible.  It is worth emphasizing here that an important advantage
of these data is that all the ellipticities are calculated within the
\textit{same} aperture ($0.3 r_{\rm vir}$).
The mean and dispersion for the sample of \citet{betal05} are
$\langle \epsilon_X \rangle = 0.376 \pm 0.019$ and
$\sigma_{\epsilon} = 0.122 \pm 0.014$, respectively
($\epsilon = 1 - \eta^2$).  The dispersion is
less than expected, but only by approximately $1\sigma$.  However, the
mean is substantially larger than expected (by approximately $2.8\sigma$).
This appears to be due to the expected mass dependence of axial ratios.

As discussed in Paper I, dark matter haloes are systematically more triaxial
the larger their mass.  A simple relation was found that describes this
behavior: $\langle s \rangle = 0.54(M_*/M_{\rm vir})^{0.05}$ (see Paper I).
Many of the clusters in the sample discussed by \citet{betal05} are much
more massive than the sample of simulation clusters we used to make our
predictions (Eq. \ref{prediction}).  This is to be expected because the
observational sample was flux limited and, therefore, massive clusters are
overrepresented (relative to a volume--limited sample, which the simulation
clusters represent).  Using the scaling relation above, we find that our
prediction for the mean ellipticity of a sample of clusters with a mass
function like that of the clusters analysed by \citet{betal05} would be
$\langle \epsilon_X \rangle = 0.353 \pm 0.013$ (instead of Eq.
\ref{prediction}).  \footnote{ We do the calculation by generating a Monte
Carlo set, picking the observationally estimated mass of a cluster, and
using the scaling relation above to get a corresponding $\langle s \rangle$.
We then draw axial ratios for the cluster using the form of the distribution
of $s$ and $q$ found in simulations (see Paper I).}
The remaining difference could well be a statistical
fluctuation, given that the intrinsic dispersion in ellipticities is
$\sigma_{\epsilon} \sim 0.14$.  Thus, we can expect fluctuations
$O(\sigma_{\epsilon}/\sqrt{N}) = 0.022$ for a sample of the size of the
\citet{betal05} sample.  By contrast, if we assume the milder scaling
advocated by JS, $\langle s \rangle = 0.54(M_*/M_{\rm vir})^{0.03}$, the
predicted mean ellipticity would be
$\langle \epsilon_X \rangle = 0.312 \pm 0.011$ instead,
which is $3.4\sigma$ lower than the observations.  It is also worth
pointing out that the data themselves show evidence of mass dependence,
although not at a high level of confidence.  If we split the data of
\citet{betal05} by mass, for clusters below (above)
$M_{\rm vir} = 10^{15} h^{-1} M_{\odot}$
$\langle \epsilon_X \rangle = 0.357 \pm 0.027$ and
$\sigma_{\epsilon} = 0.117 \pm 0.020$
($\langle \epsilon_X \rangle = 0.393 \pm 0.026$ and
$\sigma_{\epsilon} = 0.123 \pm 0.019$).  There are about equal number
of clusters in each subsample.  Although the difference in mean value
is not highly significant, it is of the magnitude expected (10\%) using
the scaling of Paper I.

We conclude from the comparison with these four data samples
that the predictions for cluster X-ray shapes in the $\Lambda$CDM
cosmology, {\it assuming gas cooling has only a small effect on the
shape of their dark matter haloes}, are in good agreement with the data.
A more stringent test, however, would require a larger sample of clusters
and a better quantitative understanding of the effect of cooling.

We have attempted to estimate quantitatively the effect
of gas cooling on cluster X-ray ellipticities, which generically makes
DM haloes less triaxial.  We use a hydrodynamic simulation of one cluster
for which cooling and star formation was abruptly terminated at redshift
$z = 2$ in order for the cluster to have reasonable star and gas fractions
\citep[see][]{ketal04}.
The effect of cooling on DM halo axial ratios for this cluster agrees
very well with the {\it average} effect shown in Figure 4 of
\citet{setal04}.  We calculated the short/long axial ratio in logarithmic
radial distance bins in order to directly compare to the figure in
\citet{setal04}.  We find that there is good agreement in the distance
range $(0.1-0.3) r_{\rm vir}$.  Therefore, we have estimated the expected
effect on X-ray ellipticities in two ways.  We can compute the change
in ellipticity by comparing the ellipticities with and without cooling
for this one cluster.  Since the change in axial ratios seems to be
representative of the expected average change, we can estimate that the
effect would be to make X-ray ellipticities 10--20\% smaller.  We have
also estimated the effect by generating a catalog of DM ``haloes'',
where a halo is represented as a set of axial ratios and a concentration.
We generate axial ratios using the form of the distribution of $s$ and
$q$ found in simulations (we use $\bar{s}=0.54$ and $\sigma_s=0.1$ for
the gaussian distribution of $s$; PaperI).  We generate concentrations
using the lognormal
distribution of \citet{wetal02}, with mean of $7$ and a log--dispersion
of $0.14$.  Finally, we orient randomly the principal axes in a box.
A mean short/long axial ratio $\bar{s}=0.54$ instead of $\bar{s}=0.45$
adequately represents the effect seen on average by \citet{setal04}, and
the effect on the cluster discussed here.  We find in this case that we
can expect  X-ray ellipticities to be $\sim 25\%$ smaller.  It is thus
rather surprising that we find the level of agreement with the data
that we have found here {\it without} taking the effect of cooling into
account.

We have considered a lower--$\sigma_8$ cosmology, in which DM halos are
predicted to be more triaxial (see Paper I), as a possible explanation
of this surprising result.  If DM halos were more triaxial, the predicted
X-ray ellipticities would increase and then cooling could bring results
into agreement with the data.  We have found in Paper I that a simple
scaling relation accounts for the dependence of axial ratios on $\sigma_8$
(see Paper I, Eq.(7)).  The predicted $\bar{s}$ can then be used as above
to generate a catalog of axial ratios.  We find that even for a value of
$\sigma_8$ as low as $\sigma_8 = 0.75$, the expected mean X-ray ellipticity
of a sample like the \citet{betal05} sample changes only to
$\langle \epsilon_X \rangle = 0.378 \pm 0.013$ (from
$\langle \epsilon_X \rangle = 0.353 \pm 0.013$ for $\sigma_8 = 0.9$).

There are potential biases that can affect comparisons of the model with
observations.
For example, in relaxed cooling flow clusters the temperature decreases
toward the center in the cluster core \citep[e.g.,][]{dgm02,vetal05a}.
Line emission of low-temperature X-ray gas can significantly alter
the $\rho_{gas}^2\sqrt{T}$ weighting assumed in our analysis and, therefore, 
the shape of the X-ray brightness isophotes.  To take this effect into
account, however, we need to know the temperature and metallicity
distribution in clusters.  We plan to address this issue in future work,
which will make use of the cluster simulations with galaxy formation.
However, \citet{betal05} have calculated ellipticites in annuli as well,
i.e. excluding the cluster centers altogether.  The mean ellipticity is
only slightly (4\%) higher.

Another possible source of bias can arise in comparisons with the
shape estimates based on the isophotes defined at a constant fraction
of the peak flux of the cluster. The profiles of real clusters are often
quite ``cuspy'' in their centers \citep[e.g.,][]{vetal05b}
and are considerably steeper than the radial gas density profiles of
clusters in our adiabatic simulations. This difference in the radial
gas distribution will result in different radii of the
isophotes defined with respect to the peak flux. This may mean that
the shapes would be measured at systematically different radii in
simulations and observations (smaller radius in observations).
Note, however, that \citet{betal05} calculate ellipticities within
the same apperture, as we have stressed above.

Two qualitative trends in X-ray maps appear to reflect the more complex
nature of dark matter haloes seen in high resolution simulations.
\citet[][and references therein]{bc96} have pioneered detailed studies of
X-ray maps to constrain cluster halo profiles. They studied 5 Abell clusters
using oblate and prolate spheroids in order to bracket the possibilities,
and concluded that the ellipticity (there is only one axial ratio if one
assumes oblate and prolate spheroids) of the haloes was constrained to be
in the range $0.40-0.55$. The systematic trend of interest here is that
4 of the 5 clusters show a decreasing ellipticity of the X-ray isophotes
at larger radii.  A similar trend
can be seen in the gas data for the two high--resolution simulation
clusters discussed in Appendix~\ref{xray} (see Figure~\ref{LCDMSBmap}
and Figure~\ref{SCDMSBmap}) and is due to the decreasing triaxiality of the
dark matter halo at larger radius. The effect is not very pronounced,
so the simple isothermal/homeoidal 
halo model could still be used for the ellipticity comparison
above.   The same is not true, however for 
Sunyaev--Zel'dovich decrement maps (see Figure~\ref{SZEmap}).
As discussed in  Appendix~\ref{xray}, the different sensitivity of
SZ maps to density and temperature ($\propto \int \rho_{gas} T$)
make these observations more sensitive to our simplistic
assumptions, and the simple mapping from halo shape
parameters will break down more visibly.  More detailed modeling
will likely be required to interpret SZ shape measurements
accurately.

A second complication 
of interest here was noted by \citet{metal89}, who pointed
out that a fraction ($\sim 15\%$) of their clusters
exhibited isophotal twist with a ``continuous rotation of the intermediate
isophotes''. We have found that one of the high-resolution simulation
clusters discussed in Appendix~\ref{xray} (see Figure~\ref{SCDMSBmap})
shows this kind of twist due to coherent twist of the dark matter density
shells. Of course, it will be interesting to quantify the frequency of
this effect, as well as its origins. The degree of misalignment in the case
of this cluster ($\sim 45^\circ$ in the radial range $\sim(0.3-1)r_{\rm vir}$)
would be rare judging by the results of JS for 12 haloes. However, a direct
comparison is not possible because the angles involved are not the same.

Therefore, while some observational quantities are somewhat
insensitive to the complex non-homeoidal nature of halo structure,
many observed properties are quite sensitive to changing ellipticities
and twists.  Specifically, the higher-order trends in 
halo shapes may leave imprints in cluster gas that 
could be studied in detail by analyses
of X-ray and Sunyaev--Zel'dovich maps.

Clusters X-ray ellipticities can be expected to evolve with redshift
due to increased halo triaxiality (see Paper I, and references therein).
Recent papers have called attention to a possible significant
evolution of the ellipticity with redshift even over the nearby
redshift range $z=0-0.1$ (\citealt*{melott01}; \citealt{plionis02}), and
have claimed that cluster X-ray and optical profiles are a little less
flattened than predicted by dissipationless and hydrodynamic simulations
(see \citealt{floor03}; \citealt{floor04}; and references therein).
However, it is important to compare observational data to simulated
clusters of similar mass (the Floor et al. clusters were more massive
than most of the observed clusters) and, as we have explained (see
Appendix~\ref{xray}), to mimic the way the data was treated.  It is hard to
draw clear conclusions when the rather different \citet{metal89} and
\citet{ketal01} X-ray data sets and analyses are combined, as was done
by \citet{melott01}.  \citet{jetal05} have detected evolution in cluster
morphology in a more homogeneous sample of clusters with \textit{Chandra}
data.  However, the morphology is not quantified as an ellipticity,
therefore we cannot assess how well this observation constrains theory
in our current paper.

\section{Conclusions}

We have presented a simple analytic model 
to predict cluster halo gas profiles based on dark halo shapes, under
the assumption that clusters are isothermal and in hydrostatic
equilibrium within haloes that are homeoidal ellipsoids (i.e. with
constant axial ratios).  We found that certain observational
properties, such as ellipticities of X-ray surface brightness maps,
can be adequately described by this modeling.  When applied to our
sample of cluster-mass haloes we find that the predicted distribution
is in good agreement with observational samples of ellipticities for
galaxy clusters.  The agreement with the recent \citet{betal05} analysis
of a complete \textit{ROSAT} sample is especially impressive, because we
found it important to take into account the predicted mass dependence of
halo shape in comparing to this data sample rich in very massive clusters.
The usefulness of our model is sensitive to how the observed
ellipticity is defined.  Specifically, care must be taken to avoid
definitions that make the calculated ellipticities sensitive to mergers.

The shape of dark matter haloes undoubtedly cannot be fully
characterized by simple models with constant axis ratios.  While we
have used inertia-tensor derived axial ratios to characterize halo
shapes in a simple way, and explored how simple assumptions about halo
shapes can be used to compare to observational tracers of halo
structure, we find that more detailed predictions will be required for
many observational comparisons (Appendix~\ref{xray}).  For example, the
isothermal/homeoidal assumption becomes less useful for comparison to
measurements like Sunyaev--Zel'dovich decrement maps.  In addition,
radially decreasing ellipticities can arise from the changing shape of
isodensity contours with radius, and twists in X-ray isophotes can
arise from misalignment of contours at large and small radius.
Predictions aimed at this kind of data will require a more detailed
analysis of $\Lambda$CDM halo shapes, including a detailed
characterization of halo ellipticities as a function of radius, 
and the frequency of isophotal twists.  Work in this direction is under
way.

\section*{Acknowledgments}
We thank Daisuke Nagai and Andrew Zentner for their help.
JSB was supported by NSF grant AST-0507816 and by startup funds at UC
Irvine.  AVK was supported by the NSF under grants
AST-0206216 and AST-0239759, and by NASA through grant NAG5-13274. BA
and JRP were supported by AST-0205944 and NAG5-12326.  The numerical
simulations used in this study were performed at the
National Center for Supercomputing Applications (NCSA,
Urbana-Champaign). This research has made use of NASA's Astrophysics
Data System Bibliographic Services.

\appendix

\section{Gas Density Inside Triaxial Halos} \label{gas}

Here we present a simple model of the gas density expected inside a cluster
halo, and use it to calculate X-ray properties such as surface brightness.
The model can also be used for other gas--density--dependent observations,
such as SZE maps from millimeter--wave observations of clusters
\citep[see e.g.][]{chr02}.
We define the dark matter halo density model, and calculate its potential, in
Appendix~\ref{potential}. The gas density model is based on three common
approximations about the gas:\newline
1) the gas is in hydrostatic equilibrium,\newline
2) the gas is isothermal,\newline
3) the gas makes a negligible contribution to the total mass.\newline
These assumptions are quite restrictive, although it is trivial to modify
equation (\ref{rhogas}) below for a polytropic gas.
In Appendix~\ref{xray}
we shall relax all of the assumptions and work directly with the gas in
two high--resolution simulations of galaxy clusters. We work out expected
X-ray properties for the clusters in the simulations, and compare them to
the predictions based on the model described here. We find that the model
works fairly well, despite its simplifying assumptions.  We further test
the model statistically against a small sample of high--resolution
simulation clusters in Section~\ref{stats}.

With the assumptions listed above, the gas density expected at a point
$(x,y,z)$ inside a triaxial halo can be written in terms of the gas density
at the origin, the dark matter potential $\Phi(x,y,z)$, and the gas
temperature $T$. For the halo density model discussed in
Appendix~\ref{potential}, we find it convenient to work with the potential
in units of the overall factor $4\pi G s q \rho_s R_s^2\,$. Therefore, we write

\begin{equation}
\label{rhogas}
\rho_{\mathit{gas}}(x,y,z) = \rho_{\mathit{gas}}(0)\,\mathrm{exp}\left\{
-\Gamma \left( \tilde{\Phi}(x,y,z)-\tilde{\Phi}(0) \right) \right\}\,.
\end{equation}

\noindent
Here $\tilde{\Phi}(x,y,z) = \Phi(x,y,z)/4\pi G s q \rho_s R_s^2\,$, so
the constant $\Gamma$ is given by

\begin{equation}
\Gamma = 4\pi G s q \rho_s R_s^2\,\frac{\mu\,m_p}{k\,T}\,,
\end{equation}

\noindent
where $\mu$ is the mean molecular weight. For clusters with galaxy velocity
dispersion $\sigma$, $kT \sim \mu m_p\sigma^2$ (see e.g. \citealt{rbn02}).
Therefore, since
$4\pi sq\rho_s R_s^2 = \mathrm{O}(\sigma^2/G)\,c_{\mathit{\rm vir}}$ (see
Section~\ref{stats}), we can expect $\Gamma \sim c_{\mathit{\rm vir}}\,$.

We can use this simple model to calculate the expected X-ray
SB of hot gas in a dark halo with a given potential. 
Since we have assumed the gas is isothermal,
SB $\propto \int{\rho_{gas}^2}\,$, where the integral is calculated along
the LOS. In Appendix~\ref{potential}, we calculate the
potential $\Phi(x,y,z)$ in the principal--axis coordinate system of the
dark matter halo. Therefore, in order to calculate SB we need the
orientation of the LOS in this coordinate system. We use the conventions
of \citet{b85}, in which the LOS--axis is defined by azimuthal and polar
angles $\phi$ and $\theta$, respectively.

We thus have the following model to predict the X-ray SB map expected for
a given projection of a dark matter halo in a simulation box. We
first calculate the axial ratios $s < q < 1$ by the iterative procedure
described in Section~\ref{stats}; in Appendix~\ref{xray} we find that
axial ratios calculated using a solid ellipsoid of semi--major axis
$0.5r_{\rm vir}$ works well to predict flux--weighted ellipticities.
We also obtain from the procedure the orientation ($\phi$ and $\theta$)
of a given LOS, and the orientation (position angle, PA) of the projection
of the shortest axis of the halo on the plane perpendicular to the LOS.
For a given point along the LOS, we find
$\rho_{\mathit{gas}}$ by first rotating its coordinates in the plane by
the PA. We then apply the inverse of the rotation parametrized by
$\phi$ and $\theta$ \citep{b85}. This gives us the coordinates of the
point along the LOS in the principal--axis system, from which we obtain
$\rho_{\mathit{gas}}$ using equation (\ref{rhogas}). Therefore, we can
calculate $\int{\rho_{gas}^2}$ numerically at any given point on the plane.
We shall refer to this model for the SB as the ``analytic model'' (even
though it involves numerical integration), in order to distinguish its
predictions from those we work out directly from the gas density in
two high--resolution simulations of galaxy clusters that we analyse in
Appendix~\ref{xray}, where we compare predictions for X-ray ellipticities.

\section{Analytic Potential of Triaxial Generalized
NFW Halos} \label{potential}

Here we consider the potential of triaxial dark matter haloes with a
density profile that is a simple generalization of a
special case of the spherical profile introduced by \citet{h90}. We
assume that isodensity shells are homeoidal ellipsoids, i.e. with constant
axial ratios
$s$ and $q$ ($s < q < 1$), and that the density profile is given by

\begin{eqnarray}
\label{density}
\rho(x,y,z) =
\frac{\rho_s}{(R/R_s)^\alpha(1+R/R_s)^{\eta-\alpha}} \nonumber \\
\nonumber \\
R = \sqrt{x^2+y^2/q^2+z^2/s^2}
\end{eqnarray}

\noindent
where $\rho_s$ and $R_s$ are a scale density and radius, respectively.
Assuming constant axial ratios allows us to reduce the calculation of
the potential to a one--dimensional integral, which in some cases can
be solved analytically, after a simple transformation of the general
result for ellipsoidal mass distributions \citep{c1969}.

We first consider $\eta = 3$. This was found to be a good
approximation (assuming constant axial ratios) by \citet{js02} for
their 12 high--resolution haloes. We also find this to be a good
approximation for the haloes of two high--resolution hydrodynamical
simulations we discuss in Appendix~\ref{xray}. However, since
spherical fits to large samples of haloes find deviations from this
value \citep{avila_reese_etal99,tetal01}, we also generalize the
results to other values below.

The potential of a thin homeoid of mass M (axes $a > b > c$), at a point
$(x,y,z)$ outside the shell, can be written as \citep{c1969}

\begin{equation}
\label{homeoid}
\Phi_M(x,y,z) = -\frac{GM}{2}\int_{\lambda }^{\infty} \frac{du}
{\sqrt{(a^{2} + u)\,(b^{2} + u)\,(c^{2} + u)}}\,.
\end{equation}

\noindent
The parameter $\lambda$ in equation (\ref{homeoid}) is the parameter of the
confocal ellipse passing through $(x,y,z)$; it is the largest root of
\begin{equation}
\frac {x^{2}}{a^{2} + \lambda } +
\frac {y^{2}}{b^{2} + \lambda } +
\frac {z^{2}}{c^{2} + \lambda } = 1\,.
\end{equation}

\noindent
Since the integral (\ref{homeoid}) can be solved analytically,
we find that

\begin{equation}
\Phi_M(x,y,z) = -\frac{GM}{\sqrt{a^2-c^2}}\ \mathrm{EllipticF}
\left(
\sqrt{\frac{a^2-c^2}{a^2+\lambda}},\sqrt{\frac{a^2-b^2}{a^2-c^2}}\,
\right)
\end{equation}

\noindent
The potential inside the homeoid is a constant \citep{c1969}, therefore
it is given by $\Phi_M(x,y,z)$ with $\lambda = 0$.

We can construct the potential inside a triaxial NFW--type halo now,
assuming homeoidal symmetry (i.e. constant axial ratios).
First, for the potential at (x,y,z) due to all mass shells inside
(i.e. inside the shell passing through the point) we find

\begin{eqnarray}
\label{P_in}
\Phi_{\mathit{in}} = -A\,\zeta^{2-\alpha}\,
\int_0^1 dm\, \frac{m^{1-\alpha}}{(1+m\zeta)^{3-\alpha}}\times \nonumber \\
\nonumber \\
\mathrm{EllipticF} \left( \sqrt{\frac{1-s^2}{1+\lambda(m)/m^2R^2}},\,
\sqrt{\frac{1-q^2}{1-s^2}}\,\right)\,.
\end{eqnarray}

\noindent
Here $\zeta = R/R_s$, $A = 4\pi G s q \rho_s R_s^2/\sqrt{1-s^2}$,
and $\lambda(m)$ is the largest root of

\begin{equation}
\frac {x^{2}}{m^{2}\,R^{2} + \lambda } +
\frac {y^{2}}{m^{2}\,R^{2}\,q^{2} + \lambda } +
\frac {z^{2}}{m^{2}\,R^{2}\,s^{2} + \lambda } = 1\, .
\end{equation}

\noindent
Second, for the potential due to all shells outside we find

\begin{eqnarray}
\label{P_out}
\Phi_{\mathit{out}}= -\frac{A}{2-\alpha}\,\mathrm{EllipticF}
 \left( \sqrt{1 - s^{2}},\,\sqrt{\frac{1-q^2}{1-s^2}} \right)\times
\nonumber \\
\nonumber \\
\left(1-\left( \frac{\zeta}{1+\zeta} \right)^{2-\alpha} \right)\,.
\end{eqnarray}

\noindent
The total potential is then
$\Phi(x,y,z) = \Phi_{\mathit{in}}+\Phi_{\mathit{out}}\,$. Also, we find
that the only change needed for $\eta \not= 3$ is to replace
$(1+m\zeta)^{3-\alpha}$ by $(1+m\zeta)^{\eta-\alpha}$ in
equation (\ref{P_in}), and to replace $(1-(\zeta/(1+\zeta))^{2-\alpha})$ in
equation (\ref{P_out}) by

\begin{eqnarray}
\frac {(2-\alpha)\,\Gamma (2-\alpha)\,\Gamma (\eta-2)}
{\Gamma (\eta - \alpha)} - \nonumber \\
\nonumber \\
\frac {\zeta ^{2-\alpha}\,\mathrm{hypergeom}([1,\, \eta - \alpha],\,
[3-\alpha],\, \zeta/(1+\zeta))}{(1+\zeta)^{\eta - \alpha}}\,.
\end{eqnarray}

For spherical symmetry ($q = s = 1$) we can check the standard result for
the potential of an NFW halo. We can use
$\mathrm{EllipticF}(x,1) = x + \mathrm{O}(x^{3})$ to get

\begin{eqnarray}
\frac {1}{\sqrt{1-s^2}}\,\mathrm{EllipticF} \left(
\sqrt{\frac{1-s^2}{1+\lambda(m)/m^2R^2}},\,
\sqrt{\frac{1-q^2}{1-s^2}}\,\right) \nonumber \\
\nonumber \\
=\, \frac{1}{\sqrt{1+\lambda(m)/m^2R^2}}\,.
\end{eqnarray}

\noindent
In this case ($q = s = 1$) the right hand side is just $m$, and for the
NFW profile ($\alpha = 1, \eta = 3$) we have

\begin{equation}
\int_0^1 \frac {m\, dm}{(1 + m\,\zeta)^2}\,=
\frac {\mathrm{ln}(1+\zeta)+\mathrm{ln}(1+\zeta)\,\zeta - \zeta}
{\zeta^2\,(1 + \zeta)}\,.
\end{equation}

\noindent
Therefore, at radial distance $r = \sqrt{x^2+y^2+z^2}$

\begin{eqnarray}
\Phi_{\mathit{in}}(r)= - 4\,\pi \,G\,{\rho_s}\,{R_s}^2\,\left(
\frac{\mathrm{ln}(1 + r/R_s)}{r/R_s} - \frac{1}{1 + r/R_s} \right) \nonumber\\
\nonumber \\
\Phi_{\mathit{out}}(r)= - 4\,\pi \,G\,{\rho_s}\,{R_s}^2\,\left( 1-
\frac{r/R_s}{1 + r/R_s} \right)
\end{eqnarray}

\noindent
and the total potential takes the standard form,

\begin{equation}
\Phi(r)=-4\,\pi\,G\,{\rho_s}\,{R_s}^2\,\frac{\mathrm{ln}(1+r/R_s)}{r/R_s}\,.
\end{equation}

\noindent
For $\alpha \not= 1$ and $\eta \not= 3$, the integral in equation
(\ref{P_in}) can be obtained analytically. The spherical potential
of a generalized NFW halo is then

\begin{eqnarray}
\Phi(r)=-4\,\pi\,G\,{\rho_s}\,{R_s}^2\,\left\{
\frac{\Gamma (2-\alpha)\,\Gamma (\eta-2)}{\Gamma (\eta-\alpha)}\right. +
\nonumber \\
\nonumber \\
\left. \frac{(r/R_s)^{2-\alpha}\,\Delta(r/R_s)}
{(2-\alpha)(1+r/R_s)^{\eta-\alpha}} \right\}\,,
\end{eqnarray}

\noindent where
\begin{eqnarray}
\Delta(x)=\frac{2-\alpha}{3-\alpha}\,\mathrm{hypergeom}\left(
[1,\eta-\alpha],[4-\alpha],\frac{x}{1+x}\right) \nonumber \\
\nonumber \\
-\, \mathrm{hypergeom}\left([1,\eta-\alpha],[3-\alpha],\frac{x}{1+x}\right).
\end{eqnarray}

\section{A Comparison of X-ray Ellipticity Measures} \label{xray}

%%%%%%%%%%%%%%%%%%%%%%%%%%%       Table1        %%%%%%%%%%%%%%%%%%%%%%%%%
\begin{table*}
\begin{center}
\caption{Ellipticity results for $\Lambda$CDM cluster;
see Figure~\ref{LCDMSBmap}.  Here
$\epsilon_X = 1 - \Lambda_-^2 / \Lambda_+^2$ \label{tbl-1}}
\begin{tabular}{ccccccccc}
\hline
\hline
{LOS} & {$\epsilon_X^{gas}$} & {$\epsilon_X^{model}$} & {$\epsilon_X^{gas}$} &
{$\epsilon_X^{model}$} & {$\epsilon_X^{gas}$} & {$\epsilon_X^{model}$} &
{$\epsilon_X^{gas}$} & {$\epsilon_X^{model}$} \\
 & {$> 0.01$} &  {$> 0.01$} &  {$> 0.1$} & {$> 0.1$} & {$0.01-0.08$} &
{$0.01-0.08$} & {$0.1-0.2$} & {$0.1-0.2$} \\
\hline
x--axis & 0.61 & 0.48 & 0.79 & 0.49 & 0.43 & 0.46 & 0.72 & 0.53 \\
y--axis & 0.24 & 0.23 & 0.40 & 0.24 & 0.28 & 0.23 & 0.32 & 0.26 \\
z--axis & 0.65 & 0.51 & 0.77 & 0.53 & 0.51 & 0.50 & 0.69 & 0.48 \\
\hline
\end{tabular}
\end{center}
\end{table*}
%%%%%%%%%%%%%%%%%%%%%%%%%%%%%%%%%%%%%%%%%%%%%%%%%%%%%%%%%%%%%%%%%%%%%%%%%

Here we evaluate the reliability of our method for predicting individual
cluster X-ray ellipticities based on the calculated axial ratios of dark
matter haloes.
Using a hydrodynamical simulation (with cooling) of a single cluster
\citet{bt95} found that this method, assuming isothermal gas,
allows an accurate estimation of the ellipticity of the dark matter even
if the gas has a strong temperature gradient, so long as any
substructure in the cluster is excluded in the analysis.
We use high--resolution
{\it adiabatic} hydrodynamical simulations
of two clusters in order to calculate X-ray SB maps directly from the gas
data of the simulations.  We then compare these maps in detail to
predictions based on the properties of their dark matter haloes, using the
theoretical model described in Appendix~\ref{gas}.
We find that the theoretical model can
work relatively well (predicting ellipticities within 10\% of the gas--data
values) depending on exactly how the observational ellipticity is defined.
A statistical (rather than case--by--case) test of the model is presented
in Section~\ref{stats}.

\begin{figure}
\epsfig{file=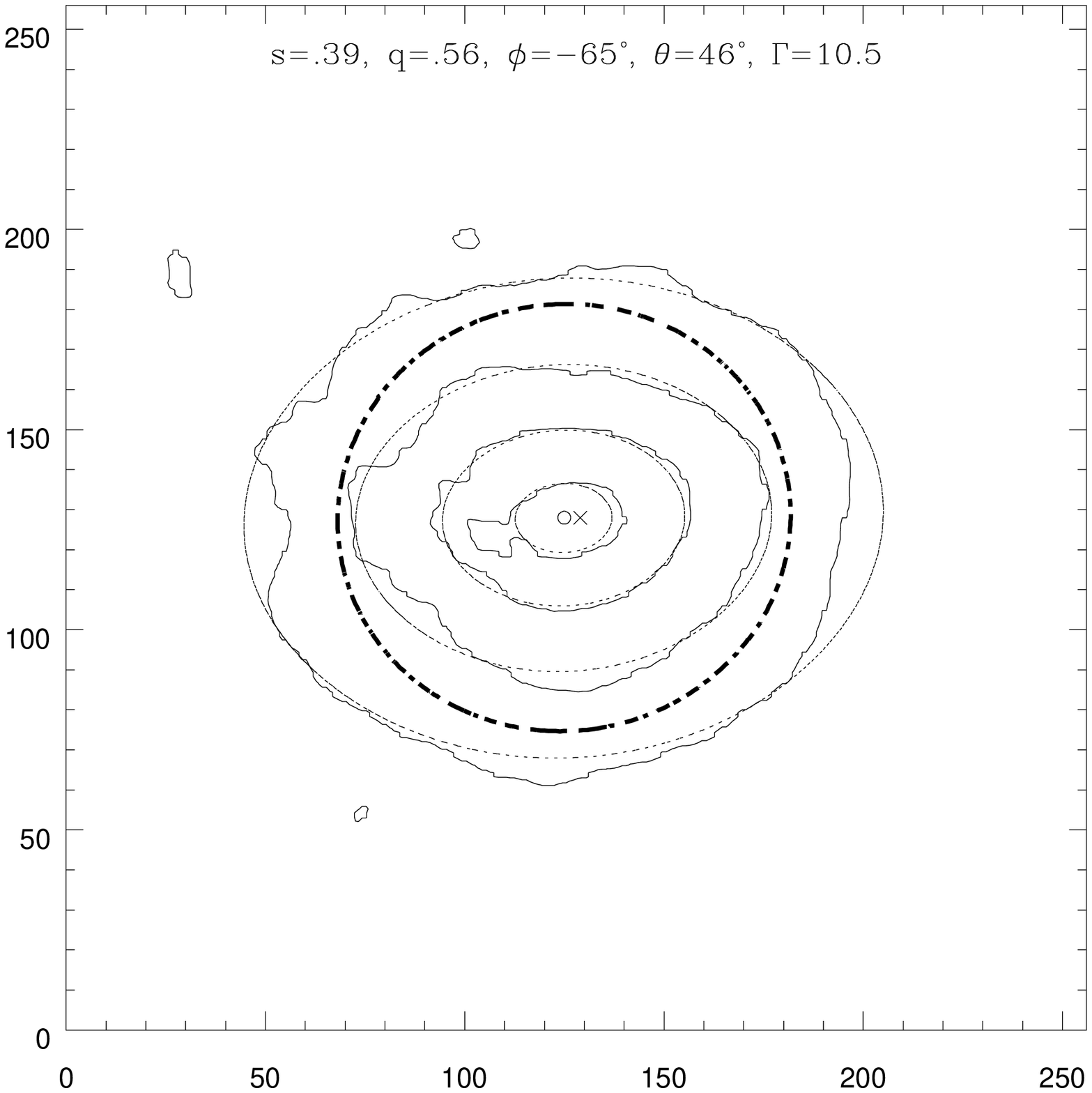,width=0.99\hsize}
\caption{Surface brightness plot for the $\Lambda$CDM cluster
in the yz--plane of the simulation box. The axes are in pixels
($7.8 \ h^{-1}$ kpc per pixel) and the solid lines show contours of constant
$\int{\rho_{gas}^2}$, spaced by factors of 10. The SB peak (centroid) is
indicated by the cross (open circle), and the innermost solid contour
corresponds to a level of 15\% of peak value. The dotted--line contours show
the predictions of the analytic model described in Appendix~\ref{gas}.  The
factor $\Gamma$ is estimated by fitting, for one projection, the radial SB
profile of the simulation. It is then used for all other projections.
Finally, the dashed--line contour illustrates a predicted contour if a more
global measure of triaxiality were used. See the text for further discussion.
\label{LCDMSBmap}}
\end{figure}

We first discuss a $\Lambda$CDM cluster that has been studied in detail by
\citet{nk03}. In Figure~\ref{LCDMSBmap} we show a ``surface brightness''
map calculated
from the gas data of the simulation. The solid lines really show contours of
constant value of $\int{\rho_{gas}^2}$, where the integration is along a
LOS parallel to the x--axis of the simulation box. Of
course, SB $\propto \int{\rho_{gas}^2\sqrt{T}}$, but we
have dropped the temperature dependence for simplicity, given that it makes
only a small difference in calculated ellipticities ($\lsim 5\%$; see
below).  We calculate SB for a given ``pixel'' by summing over all cells 
along the LOS.
Coordinates are shown in pixels, with $7.8 \ h^{-1}$ kpc per pixel. The
dotted--line contours are the SB contours predicted by the model described
in Appendix~\ref{gas} with the factor $\Gamma$ chosen to match the radial
SB profile, $\Gamma = 10.5$. For this halo $c_{\mathit{\rm vir}} = 11.5$,
therefore $\Gamma \sim c_{\mathit{\rm vir}}\,$, as expected (see
Section~\ref{stats}). The X-ray ellipticities discussed here are only mildly
dependent on $\Gamma$ (e.g. $\epsilon_X^{model}=0.46$ in Table~\ref{tbl-1}
changes to $\epsilon_X^{model}=0.40(0.48)$ for $\Gamma=8(13)$).
They are mostly sensitive to the axial ratios $s$ and $q$, and the relative
orientation of the LOS, described by polar angles $\phi$ and $\theta$ in the
principal--axis coordinate system. The axial ratios and polar angles were
calculated inside an ellipsoid of semi-major axis $600 \ h^{-1}$ kpc using
the iterative method described in the text, and are shown at the top of the
figure.  The dashed--line contour illustrates a predicted isophotal contour
based on axial ratios calculated with the prescription of \citet{ke04}.
The result is very similar for SZ isodecrement contours.

%%%%%%%%%%%%%%%%%%%%%%%%%%%        Figure7        %%%%%%%%%%%%%%%%%%%%%%%
%%%%%%%%%%%%%%%%%%%%%%%%%%%%%%%%%%%%%%%%%%%%%%%%%%%%%%%%%%%%%%%%%%%%%%%%%

The highly irregular, innermost solid contour is due to a minor merger
nearly in the ``plane of the sky'' (about $25^{\circ}$ off the y--axis
of the box).  The merger is ideal to test quantitatively observational
strategies to calculate an ellipticity that best represents the global
triaxiality of the dark matter halo. It is also an ideal test of the
reliability of our method because it allows us to gauge what bias can
be introduced in the calculation of ellipticities by the presence of a
minor merger, which can be expected to be common for cluster-size
systems. We discuss both of these issues below.

There are various strategies to calculate X-ray ellipticities. Here we
consider the method used by \citet{ketal01} (22 clusters; ROSAT data)
and \citet{metal89} (49 clusters; {\it Einstein} data) as examples used
in analyses of samples of clusters. Both studies use the method of
\citet{cm80} adapted to an X-ray image. The ellipticity is calculated from
the positive roots $\Lambda_+$ and $\Lambda_-$ ($\Lambda_+ > \Lambda_-$)
of the characteristic equation

\begin{equation}
(\mu_{20} - \Lambda^2)(\mu_{02} - \Lambda^2) = \mu_{11}^2\ .
\end{equation}

\noindent
The moments $\mu_{mn}$ are defined in terms of the flux $f_{ij}$ at a given
pixel $(x_i,y_j)$ by

\begin{equation}
\mu_{mn} = \sum_{ij} f_{ij}(x_i-\bar{x})^m(y_j-\bar{y})^n / \sum_{ij} f_{ij}\ ,
\end{equation}

\noindent
where ($\bar{x}$,$\bar{y}$) is the image centroid ($\bar{x} =
\sum_{ij} x_if_{ij} / \sum_{ij} f_{ij}$, $\bar{y} =
\sum_{ij} y_jf_{ij} / \sum_{ij} f_{ij}$). The ellipticity is then calculated as

\begin{equation}
\label{ex:metal89}
\epsilon_X = 1 - \Lambda_-^2 / \Lambda_+^2
\end{equation}

\noindent
by \citet{metal89}, and as

\begin{equation}
\label{ex:ketal01}
\epsilon_X = 1 - \Lambda_- / \Lambda_+
\end{equation}

\noindent
by \citet{ketal01}. We shall use equation (\ref{ex:metal89}) here, except
when comparing directly with the data of \citet{ketal01}.

The X-ray ellipticity $\epsilon_X$ is rather sensitive to what pixels are
used to calculate it. \citet{ketal01} use all pixels above a flux threshold
(which is the average flux within a region of given radius).  For example,
in Figure~\ref{LCDMSBmap} this threshold is $\sim 0.01$ of the peak flux
within $600 \ h^{-1}$ kpc (which is the largest radius they use to define
the threshold).  In Table~\ref{tbl-1} (columns 2--5)
we show results for the cluster of Figure~\ref{LCDMSBmap} for two flux
thresholds (0.01 and 0.1 of the peak flux)
and for a LOS along each of the axes of the simulation box.
This choice of pixels emphasizes the brightness peaks and, therefore, is more
sensitive to mergers. Thus, the analytic model prediction for the ellipticity,
$\epsilon_X^{model}$, deviates significantly from the value calculated
directly from the gas, $\epsilon_X^{gas}$,
except when the merger is nearly along the LOS. On the other
hand, \citet{metal89} explicitly exclude the centre of an image in order to
characterize the global dynamics of a cluster. They use all the pixels
containing 20\% of the flux above a faint threshold.
The latter varies substantially across the sample, but for 80\% of clusters
it is $\sim 0.01-0.2$ of the peak flux. In Table~\ref{tbl-1} (columns 6--9)
we show results for two flux ranges (0.01--0.08 and 0.1--0.2)
and for a LOS along each of the axes.
In this case the analytic model performs much better, provided the fainter
threshold is chosen low enough. If we calculate the deviation of the model
ellipticity from the gas ellipticity (using a flux threshold of $0.01$) for
$100$ random LOS we find

\noindent
(1) For the \citet{metal89} ellipticity the analytic 
model works fairly well; 2/3 of
the time the model predicts the ellipticity within 10\% of the gas value.
Also, it would not bias a statistical sample because it predicts larger and
smaller values with equal frequency.

\noindent
(2) For the \citet{ketal01} ellipticity the analytic 
model predicts a systematically
smaller value. This is expected in this case because the model misses the
merger, therefore it predicts rounder SB contours from all viewing angles.
In this case we find that 2/3 of the time the value is $20-30\%$ smaller.

\noindent
We find similar trends for the mean and the dispersion of ellipticities
calculated for the sample of clusters discussed in Section~\ref{stats},
although individual values can deviate more than indicated here.

%%%%%%%%%%%%%%%%%%%%%%%%%%%       Table2        %%%%%%%%%%%%%%%%%%%%%%%%%
\begin{table*}
\begin{center}
\caption{Ellipticity results for SCDM cluster; see Figure~\ref{SCDMSBmap}.
Here $\epsilon_X = 1 - \Lambda_-^2 / \Lambda_+^2$ \label{tbl-2}}
\begin{tabular}{ccccccccc}
\hline
\hline

{LOS} & {$\epsilon_X^{gas}$} & {$\epsilon_X^{model}$} & {$\epsilon_X^{gas}$} &
{$\epsilon_X^{model}$} & {$\epsilon_X^{gas}$} & {$\epsilon_X^{model}$} &
{$\epsilon_X^{gas}$} & {$\epsilon_X^{model}$} \\
 & {$> 0.01$} &  {$> 0.01$} &  {$> 0.1$} & {$> 0.1$} & {$0.01-0.09$} &
{$0.01-0.09$} & {$0.1-0.2$} & {$0.1-0.2$}\\
\hline
x--axis & 0.30 & 0.35 & 0.38 & 0.38 & 0.25 & 0.34 & 0.38 & 0.34 \\
y--axis & 0.48 & 0.37 & 0.45 & 0.42 & 0.51 & 0.35 & 0.42 & 0.39 \\
z--axis & 0.48 & 0.55 & 0.52 & 0.56 & 0.44 & 0.54 & 0.48 & 0.55 \\
\hline
\end{tabular}
\end{center}
\end{table*}
%%%%%%%%%%%%%%%%%%%%%%%%%%%%%%%%%%%%%%%%%%%%%%%%%%%%%%%%%%%%%%%%%%%%%%%%%

The flux level at the outermost contour in Figure~\ref{LCDMSBmap} is
$\sim 0.002$ of the peak flux. At this flux level the contour is clearly
rounder than the model prediction (due to the fact that the dark matter
halo gets rounder farther out, whereas the analytic 
model assumes constant axial ratios).
However, pixels up to much higher flux
levels ($\sim 0.06$ of peak flux) enter the calculation in order to
accumulate 20\% of the flux above this fainter threshold
in the approach of \citet{metal89}. For example, for
the x--axis $\epsilon_X^{gas} = 0.41$ and $\epsilon_X^{model} = 0.45$ in
the flux range $0.002 - 0.06$ of peak flux. Therefore, the model works well
down to lower thresholds.

The analytic model assumes that the gas is isothermal in order to predict
the SB. We can check how much this is likely to affect a comparison with
actual data by calculating the ellipticity from the simulation data including
the temperature dependence. We find that for the flux levels considered here,
the effect is rather small. For example, the entry in column 2 of
Table~\ref{tbl-1} would be 0.59 (0.23, 0.63) for the x--axis (y--axis,
z--axis) as compared to 0.61 (0.24, 0.65) assuming isothermality.
The temperature in this cluster falls by a factor $\sim 1.9$ in
the radial range $(0.1-0.5)r_{\mathit{\rm vir}}$, which is consistent with
observations \citep[see][and references therein]{dgm02}. Therefore, the
temperature variation of the simulation gas is representative of that of
real clusters.  

We have also tested whether using the dark matter potential of this
cluster directly would significantly improve the prediction for
$\epsilon_X$. The assumptions are still the same, but the potential is
calculated directly from the dark matter distribution in order to
predict the gas density.  We find that the results improve as follows.
For example, the entry in column 3 of Table~\ref{tbl-1} would be 0.53
(0.25, 0.55) for the x--axis (y--axis, z--axis) instead of 0.48 (0.23,
0.51).

\begin{figure}
\epsfig{file=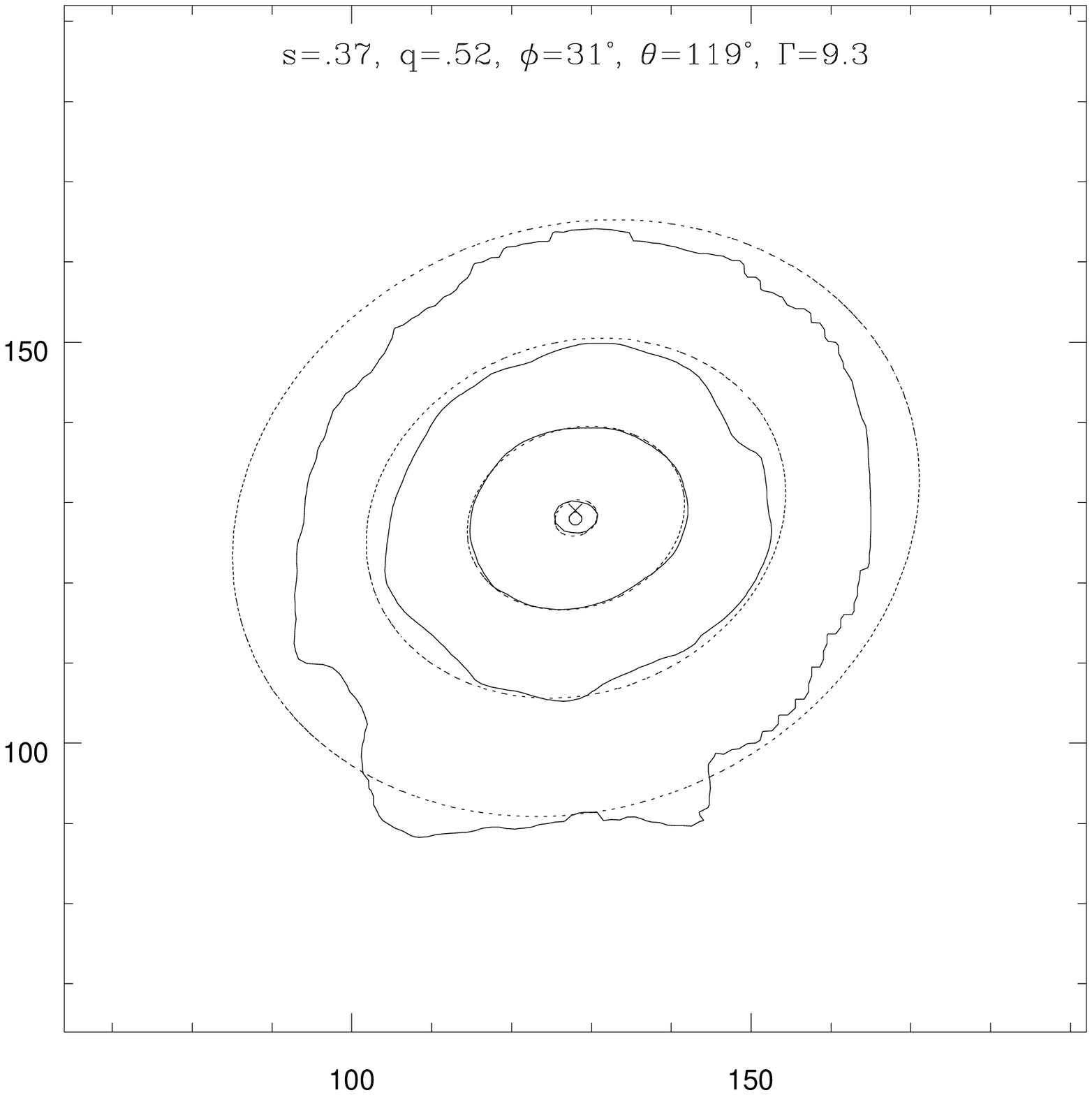,width=0.99\hsize}
\caption{Surface brightness plot for the SCDM cluster
in the yz--plane of the simulation box. The axes are in pixels
($15.6 \ h^{-1}$ kpc per pixel) and the meaning of symbols is as in
Figure~\ref{LCDMSBmap}. In this case the innermost solid contour corresponds
to a level of 25\% of peak value, and each solid contour is drawn at one
tenth of the solid--contour levels of Figure~\ref{LCDMSBmap}.
\label{SCDMSBmap}}
\end{figure}

Finally, in order to study whether the analytic model indeed performs
better in the absence of a merger, we have analysed in the same manner a
high--resolution simulation cluster that does not have an ongoing merger.
It is a SCDM cluster discussed in detail by
\citet{kkh02}. In Figure~\ref{SCDMSBmap} we show the SB map calculated as
in Figure~\ref{LCDMSBmap}, and in Table~\ref{tbl-2} we show the results for
the ellipticity. For this cluster $\Gamma = 9.3$ and
$c_{\mathit{\rm vir}} = 10.4$, therefore $\Gamma \sim c_{\mathit{\rm vir}}\,$
as before. In this case the model works reasonably well for either
one of the definitions of ellipticity, provided that the flux threshold is
sufficiently
high. For faint thresholds the model fails to reproduce the trend of rounder
and twisted SB contours in the simulation (for LOS = y--axis, the reverse
trend in Table~\ref{tbl-2} is due to the chance projection of a distant hot
spot that appears only at a level $\sim 0.01$ of peak flux). It is in fact
the twisted SB contours that cause most of the difference between model and
simulation gas. This is due to the fact that isodensity shells are fairly
misaligned in this case. The projected, $100 \ h^{-1}$ kpc--thick isodensity
shell of $400 \ h^{-1}$ kpc ($800-900 \ h^{-1}$ kpc, $1000-1050 \ h^{-1}$ kpc)
semi-major axis makes a $15^\circ$ ($35^\circ$, $60^\circ$) angle with
the vertical direction in Figure~\ref{SCDMSBmap}. Such large misalignments
were found to be rare by \citet{js02}, therefore we assume here that the
model also works down to the faint threshold level of $\sim 0.01$ of peak
flux in the absence of a merger, and for both ellipticities.

%%%%%%%%%%%%%%%%%%%%%%%%%%%        Figure8        %%%%%%%%%%%%%%%%%%%%%%%
%%%%%%%%%%%%%%%%%%%%%%%%%%%%%%%%%%%%%%%%%%%%%%%%%%%%%%%%%%%%%%%%%%%%%%%%%

In Section~\ref{clusters} we analyse the expected distribution of X-ray
ellipticities for cluster--mass haloes in the cosmological box discussed in
Section~\ref{stats}. We calculate ellipticities using the analytic model,
and compare the distribution to the  data of \citet{metal89}
and \citet{ketal01}.

\begin{figure}
\epsfig{file=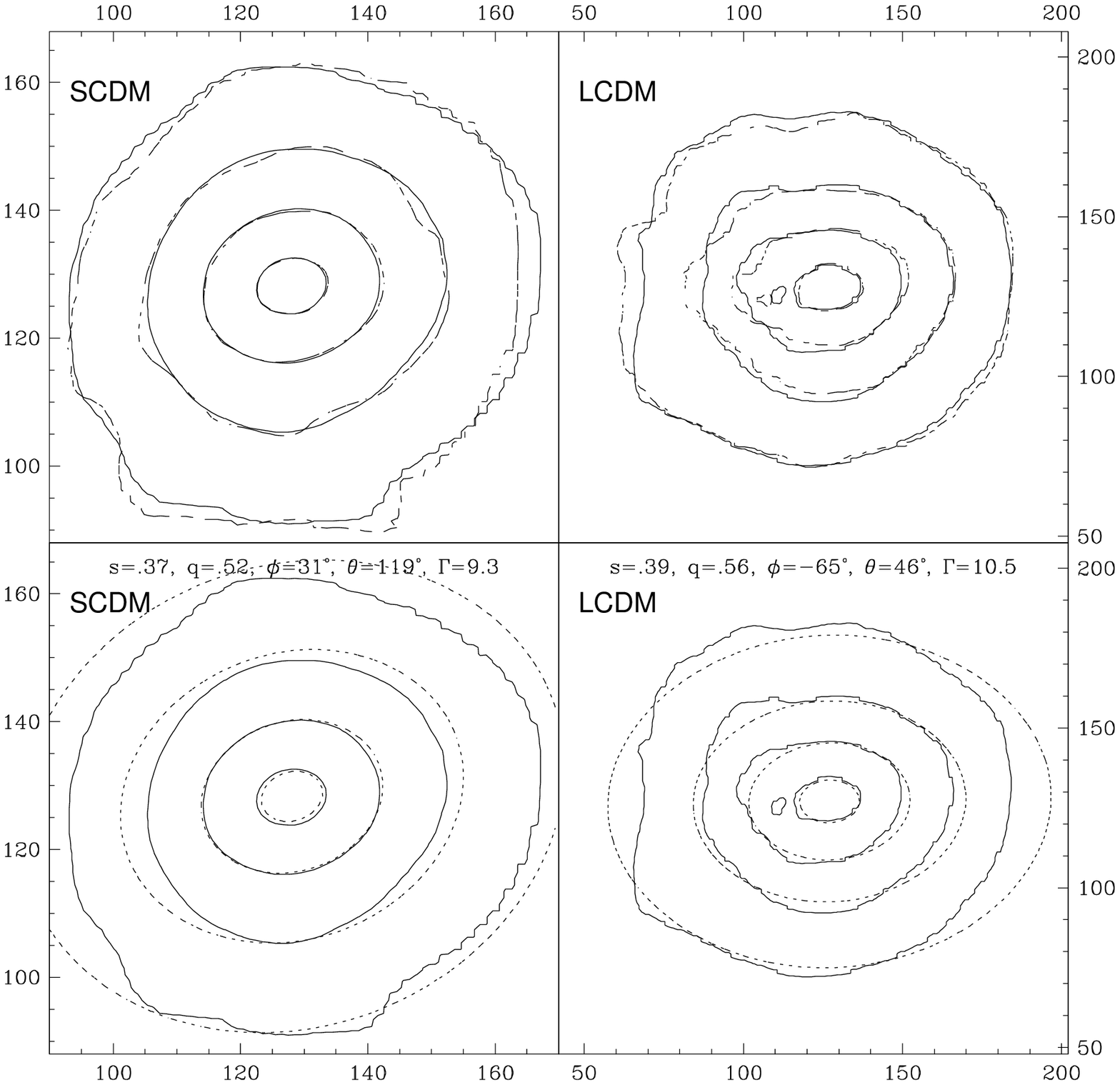,width=0.99\hsize}
\caption{Temperature decrement map for the two clusters of
Figures~\ref{LCDMSBmap} and \ref{SCDMSBmap}, in the yz--plane of the
corresponding simulation boxes. The axes are in pixels, and
each box is $1.25 \ h^{-1}$ Mpc across. The solid lines show contours of
constant $\int{\rho_{gas}\,T}$, spaced by factors of 3. The innermost
solid contour corresponds to a level of 60\% (50\%) of peak value for the
SCDM ($\Lambda$CDM) cluster. The top panels compare the shape of the contours of
constant $\int{\rho_{gas}}$ (dashed lines) to the decrement-level contours.
The bottom panels compare the prediction of the analytic model for
$\int{\rho_{gas}}$ (dotted lines) with the decrement-level contours.
See the text for further discussion.
\label{SZEmap}}
\end{figure}

We have also considered the reliability of the analytic model to
predict the shape of SZE maps of clusters. In Figure~\ref{SZEmap} we show
``temperature decrement'' maps for the two clusters we have
discussed. The solid lines show contours of constant value of
$\int{\rho_{gas}\,T}$ (spaced by factors of 3). The integration is
along a LOS parallel to the x--axis of the corresponding simulation
box, as was the case in Figures~\ref{LCDMSBmap} and
\ref{SCDMSBmap}.  Since the dependence on gas temperature is linear in
this case, we can expect a more significant effect of temperature on
the shape of isodecrement contours. The top panels of Figure~\ref{SZEmap}
compare the shape of the contours of constant value of $\int{\rho_{gas}}$
only (dashed lines) to decrement contours. \footnote{The dashed--line
contours are not shown spaced by a fixed factor.  The levels are just
chosen to give contours of similar size to the solid contours, in order
to compare shapes at a given radial distance.}  Both
sets of solid contours are calculated directly from the gas and
temperature data of the corresponding simulation. It can be seen that
in the presence of a merger (the $\Lambda$CDM cluster case) the
temperature dependence indeed makes the isodecrement contours
noticeably different from contours of $\int{\rho_{gas}}$. However, in
the absence of a merger (the SCDM cluster case) they agree fairly well
in shape. For example, the ellipticity $\epsilon_{SZE}=1-\Lambda_-^2 /
\Lambda_+^2$, calculated using the signal between the second and third
contours, is $\epsilon_{SZE} = 0.20$ ($0.24$) for the SCDM
($\Lambda$CDM) cluster. The ellipticities calculated using the gas
density alone are $0.21$ and $0.36$, respectively.  Thus, analytic models
to calculate $\epsilon_{SZE}$ assuming isothermal gas will err by a large
margin in the presence of a merger, \textit{even if $\epsilon_{SZE}$ is
calculated outside the core region}.  This is unlike what we have found
for X-ray ellipticities.

%%%%%%%%%%%%%%%%%%%%%%%%%%%        Figure9        %%%%%%%%%%%%%%%%%%%%%%%
%%%%%%%%%%%%%%%%%%%%%%%%%%%%%%%%%%%%%%%%%%%%%%%%%%%%%%%%%%%%%%%%%%%%%%%%%

Furthermore, even in the absence of a merger, the changing triaxiality of
the dark matter halo makes model predictions for ellipticity in the SZ
maps miss the values $\epsilon_{SZE}$ by a larger margin than in the case
of X-ray ellipticity.  The bottom panels of Figure~\ref{SZEmap} show
the predictions of the analytic model for $\int{\rho_{gas}}$ (dotted
lines) compared to the ``isodecrement contours'' of the top panels (solid
lines).  It can be seen there that, even in the absence of a merger (left),
the ellipticity of the analytic model contours is too large (even if compared
to the simulation--data contours for $\int{\rho_{gas}}$ only (dashed--lines
of top panels).  For example, the ellipticity between the second and third
contours of the analytic model predictions is $0.33$ ($0.46$) for the SCDM
($\Lambda$CDM) cluster. We find similar results for the other LOS. Thus,
reliable predictions (i.e. within 10\% of gas--data values) for ellipticity
in SZE maps need to incorporate the changing triaxiality of the dark matter
haloes.  However, the model is still useful to predict statistics of cluster
samples such as the mean and dispersion. See Section~\ref{stats}.


\end{document}